\documentclass[preprint,aps,nofootinbib,tightenlines]{revtex4}

\usepackage{graphicx}

\usepackage{amsmath}
\usepackage{amsfonts}
\usepackage{amssymb}

\def\beq{\begin{equation}}
\def\eeq{\end{equation}}
\def\bea{\begin{eqnarray}}
\def\eea{\end{eqnarray}}

\def\slashchar#1{\setbox0=\hbox{$#1$}           
   \dimen0=\wd0                                 
   \setbox1=\hbox{/} \dimen1=\wd1               
   \ifdim\dimen0>\dimen1                        
      \rlap{\hbox to \dimen0{\hfil/\hfil}}      
      #1                                        
   \else                                        
      \rlap{\hbox to \dimen1{\hfil$#1$\hfil}}   
      /                                         
   \fi}

\begin{document}


\vspace*{1cm}
\title{Theoretical Constraints on the Higgs Effective Couplings}

\author{\vspace{1cm}Ian Low$^{ a, b}$, Riccardo Rattazzi$^{c}$, and Alessandro Vichi$^c$}

\affiliation{\vspace{0.5cm}
\mbox{$^a$High Energy Physics Division, Argonne National Laboratory, Argonne, IL 60439, USA}
\vspace{0.2cm}\\
\mbox{$^b$Department of Physics and Astronomy, Northwestern University, Evanston, IL 60208, USA}\vspace{0.2cm}\\
\mbox{$^c$Institut de Th\'eorie des Ph\'enom\`enes Physiques, EPFL, CH-1015 Lausanne, Switzerland}\vspace{1cm}}

\begin{abstract}
\vspace*{0.5cm} 
We derive constraints on the sign of couplings in an effective Higgs Lagrangian using prime principles such as the naturalness principle,
global symmetries, and unitarity. Specifically, we study four dimension-six operators, ${\cal O}_H$, ${\cal O}_y$, ${\cal O}_g$, and ${\cal O}_\gamma$, 
which contribute to the production and decay of the Higgs boson at the Large Hadron Collider (LHC), among other things. Assuming the Higgs is a fundamental scalar, we find: 1) the coefficient of ${\cal O}_H$ is positive except when there are triplet scalars, resulting in a reduction in the Higgs on-shell coupling from their standard model (SM) expectations if no other operators contribute, 2) the linear combination of ${\cal O}_H$ and ${\cal O}_y$ controlling the overall Higgs coupling to fermion is always reduced, 3) the sign of ${\cal O}_g$ induced by a new colored fermion is such that it interferes destructively with the SM top contribution in the gluon fusion production of the Higgs, if the new fermion cancels the top quadratic divergence in the Higgs mass,
and 4) the correlation between naturalness and the sign of ${\cal O}_\gamma$ is similar to that of ${\cal O}_g$, when there is a new set of heavy electroweak gauge bosons. Next considering a composite scalar for the Higgs, we find 
the reduction in the on-shell Higgs couplings persists. If further assuming a collective breaking mechanism as in little Higgs theories, the coefficient of ${\cal O}_H$ remains positive even in the presence of triplet scalars. In the end, we conclude that the gluon fusion production of the Higgs boson is reduced from the SM rate in all composite Higgs models. Our study suggests a wealth of information could be revealed by precise measurements of the Higgs couplings, providing strong motivations for both improving on measurements at the LHC and building a precision machine such as the linear collider.
\end{abstract}


\maketitle

\section{Introduction}
\label{sec:introduction}

The coming experiments at the Large Hadron Collider (LHC), by directly exploring physics at the weak scale, are set to unveil the precise dynamics of electroweak symmetry breaking. Indeed, the accurate measurements performed at lower energies by LEP1/SLC and LEP2 already provide important indirect information. It is fair to say that, according to these results, the favored scenario is one in which the dynamics is weak  up at least the multi-TeV range and a light Higgs degree of freedom exists \cite{electroweak}. In recent years interesting theoretical progress, based on extra-dimensional constructions \cite{Csaki:2003dt},
was also made towards the construction of Higgsless theories of electroweak symmetry breaking. Constructions along those lines have  perhaps a better chance than ordinary technicolor to satisfy the electroweak precision tests. But, as far as we know right now, they compare less well to models featuring a light Higgs. In the end approaches utilizing a Higgs-like scalar appear more promising.

Quantum mechanically, a scalar particle like the Higgs receives corrections to its mass-squared that are quadratically
sensitive to ultraviolet (UV) physics, and much of the model-building activity centers on devising ways to reduce this UV 
sensitivity. On symmetry grounds two general classes of models exist to date: those using spontaneously
broken accidental global symmetries, under which the Higgs may arise as a pseudo-Nambu-Goldstone boson 
(PNGB) \cite{Kaplan:1983fs,Kaplan:1983sm}, and those
possessing low-scale supersymmetry. While supersymmetry has been the leading candidate for weak  scale physics for
several decades, the idea that Higgs may be
a PNGB was first considered many years ago by Refs.~\cite{Kaplan:1983fs,Kaplan:1983sm} and resurrected after the proposal
of little Higgs theories \cite{Arkani-Hamed:2001nc,Arkani-Hamed:2002qx,Arkani-Hamed:2002qy} and of holographic Goldstone models 
\cite{Contino:2003ve,Agashe:2004rs}. In this class of theories, the
Higgs boson arises as a composite particle out of some unknown strong dynamics at a  scale $f$. The presence of a moderate separation between the weak scale $G_F^{-1/2}\equiv v$ and   $f$ allows to keep under control the unwanted corrections to electroweak precision observables. Little Higgs models aim at explaining the smallness  of $v^2/f^2$ in a fully natural way, as a loop effect.  Holographic Goldstone models, instead, content themselves by explaining  $v^2/f^2\ll 1$ as the result  of an accidental cancellation,
which does not seem implausible given that a mild hierarchy $v^2/f^2< 0.1-0.3$ is sufficient in the existing models. Moreover, also thanks to their somewhat less ambitious goal, holographic models can rely on a simpler
structure (symmetries, couplings, and multiplets) than the ambitious little Higgs models.  Thus it is fair to keep both classes of models under consideration.
 On the other hand, there are  models, supersymmetric or not, which do not address the issue of cancellations of quadratic divergences in the Higgs mass, such as those in 
Refs.~\cite{Appelquist:2000nn,ArkaniHamed:2004fb}. In particular in the model of  Ref.~\cite{ArkaniHamed:2004fb} the presence of new states at the weak scale is associated with the dark matter problem and the gauge unification constraints, rather than with electroweak symmetry breaking.

If the Higgs boson is in the  hundred GeV range, 
and any new particles interacting with it are 
 heavier, then it is possible to integrate out the heavy states and study properties of the Higgs boson by an effective Lagrangian technique \cite{Weinberg:1978kz}. 
Different approaches to an effective Higgs Lagrangian have been followed in the literature. A first approach consists in writing down all possible operators in a large mass expansion, without making any dynamical assumption. A second, complementary, approach consists of focusing on
one very specific model, computing the effective coefficients, and scanning through the allowed parameter space. Besides these two approaches, it is possible to have a third approach instead, which corresponds roughly to marrying the  first two by making simple dynamical assumptions that encompass a general class of models.
A specific example of this third approach is given by the strongly interacting light Higgs effective Lagrangian (SILH) \cite{Giudice:2007fh}, which concentrates on the general scenario  where the Higgs is a light composite PNGB, while the rest of the SM particles are elementary. Which of the above approaches to choose depends on the questions one wants to ask.
 Now, given the uncertainty of what could (and could not) be seen at the LHC, and the realization that several
models could lead to similar first-order predictions at the LHC, we think it is important to ask structural and global
questions such as:  Is
physics at the electroweak scale natural? Is the Higgs boson fundamental or composite? If new particles are discovered, 
who ordered them? Are they responsible for canceling the Higgs quadratic divergences? In this regard, neither of the first two 
methods is adequate. For instance, in the first approach there is no assessment as to which effects are genuinely associated with the electroweak breaking dynamics and which are not. On the other hand,
the third method, based on definite dynamical assumptions, seems
better suited to ask questions which are relevant to structure and dynamics.

In this paper we initiate a study to address the aforementioned structural and global questions. We adopt a top-down approach by 
exploring possible theoretical constraints on the sign of the Higgs effective couplings from the naturalness principle and from other prime 
principles such as global symmetry patterns and unitarity. Even though we will base our work on the SILH Lagrangian in
Ref.~\cite{Giudice:2007fh}, most of our results do not depend on the SILH assumption on the PNGB nature of the Higgs, and can be applied to the case of a fundamental Higgs scalar, in particular to supersymmetric models.
The advantage of using the SILH Lagrangian is power counting: we will
be able to understand which operators are genuinely sensitive to the underlying strong dynamics occurring at the scale $f$, as well as the relative importance of different operators when it comes to the collider phenomenology.

According to the results in Ref.~\cite{Giudice:2007fh}, the most relevant effects of a strongly-interacting light Higgs
 are described by the following dimension-six  effective operators:
\begin{equation}
\label{allope}
\begin{array}{ll}
 {\cal O}_H= \partial^\mu (H^\dagger H)\partial_\mu(H^\dagger H) \quad& {\cal O}_y = H^\dagger H \bar{f}_L H f_R +{\rm h.c.} \\
  {\cal O}_g = H^\dagger H G_{\mu\nu}^a G^{a\, \mu\nu} \quad & {\cal O}_\gamma = H^\dagger H B_{\mu\nu} B^{\mu\nu} 
\end{array}
\end{equation}
where we use the same notation as in \cite{Giudice:2007fh}.
The operator ${\cal O}_H$ renormalizes the Higgs kinetic term, after the Higgs gets a vacuum expectation value (VEV)
$\langle H\rangle = (0, v/\sqrt{2})^T$,
and contributes to the high energy limit of the scattering amplitudes for longitudinal gauge bosons, while the operator
${\cal O}_y$ modifies the Higgs coupling to the SM fermions, in particular the top quark. Therefore these four operators
control, among other things, the Higgs production rate in the gluon fusion 
channel and decay width in the di-photon channel \cite{Giudice:2007fh}. 
The gluon fusion channel is the dominant production mechanism of the
Higgs at the LHC, while the decay to two photons is the main discovery mode for a Higgs mass below 140 GeV. 

This paper is organized as follows: in Section \ref{sect:powercount} we recall basic results for the SILH Lagrangian.
  Section III discusses
theoretical constraints on the sign of the coefficient of ${\cal O}_H$, followed by 
Section IV which considers constraints on ${\cal O}_g$, ${\cal O}_y$, and ${\cal O}_\gamma$. In Section \ref{sect:discussion}
we present a synthesis of our results and summarize.
 

\section{The SILH Effective Lagrangian}
\label{sect:powercount}
Many of our results are quite general and apply to theories with either a fundamental or composite Higgs scalar. Nevertheless, it will be convenient to adopt the composite Higgs model as a benchmark scenario and use the SILH Lagrangian as
a reference frame, whose results we briefly review here.

The generic scenario of Higgs compositeness to which little Higgs and holographic Higgs models
belong can be characterized as follows:
\begin{enumerate}
\item The Higgs doublet $H$ results from some new (strong) dynamics  broadly described by two parameters $m_\rho$, the mass scale of composite states,  and $g_\rho$, their self-coupling.  From this definition it follows that non-renormalizable Higgs  self-interactions are characterized by the coupling scale $f\equiv m_\rho/g_\rho$. In models where the Higgs is a PNGB the scale $f$ corresponds to the decay constant of the non-linear sigma model (nl$\sigma$m).
\item The vector bosons and the fermions of the SM are elementary up to well above the Fermi scale. The vectors are coupled 
to the strong sector via a weak gauging of a $SU(3)_c\times SU(2)_L\times U(1)_Y$ subgroup of its global symmetry group $G$. The fermions are coupled via the same interactions that give rise to Yukawa couplings. In particular, apart from the top quark, the SM fermions  are very weakly coupled to the strong dynamics.
\end{enumerate}
Under these assumptions (and with a two more specifications we shall explain below) the deviations from the SM at energy $E<m_\rho$ are described at leading order by the following dimension-six effective Lagrangian \cite{Giudice:2007fh}\footnote{This is the subset of the list in Ref.~\cite{Buchmuller:1985jz} which is compatible with assumptions 1 and 2.}:
\begin{eqnarray}
\label{L_SILH}
 {\cal L}_{\rm SILH} &=& \frac{c_H}{2f^2} \partial^\mu (H^\dagger H)\partial_\mu(H^\dagger H) + \frac{c_T}{2f^2}
 \left( H^\dagger \tensor{D}^\mu H\right)  \left( H^\dagger \tensor{D}_\mu H\right)\nonumber \\
 &&  -\frac{c_6 \lambda}{f^2} (H^\dagger H)^3 
 + \left(\frac{c_y y_f}{f^2} H^\dagger H \bar{f}_L H f_R + {\rm h.c.}\right)  \nonumber\\
&&  + \frac{i c_W g}{2 m_\rho^2} \left( H^\dagger \sigma^a \tensor{D}_\mu H\right)(D_\nu W^{\mu\nu})^a
+    \frac{i c_B g'}{2 m_\rho^2} \left( H^\dagger  \tensor{D}_\mu H \right)(\partial_\nu B^{\mu\nu}) \nonumber \\
&& +  \frac{i c_{HW} g}{16\pi^2 f^2} (D^\mu H)^\dagger \sigma^a (D^\nu H) W^a_{\mu\nu}
 +\frac{i c_{HB} g'}{16\pi^2 f^2} (D^\mu H)^\dagger  (D^\nu H) B_{\mu\nu}\nonumber\\ 
 && +\frac{c_\gamma g^{\prime\, 2}}{16\pi^2 f^2}\frac{g^2}{g_\rho^2} H^\dagger H B_{\mu\nu} B^{\mu\nu}
  + \frac{c_g g_s^2}{16\pi^2 f^2} \frac{y_t^2}{g_\rho^2}  H^\dagger H G_{\mu\nu} G^{\mu\nu} ,
\end{eqnarray}
where $g_s$, $g$, and $g'$ are the SM $SU(3)_c\times SU(2)_w\times U(1)_Y$ gauge couplings, respectively, and $y_t$ is the SM top Yukawa
coupling. Our notation and definition is such that
\begin{equation}
H = \frac1{\sqrt{2}} \left(
\begin{array}{c}
h^+ \\
h^0
\end{array} \right) =
 \frac1{\sqrt{2}} \left(
\begin{array}{c}
h^1+ i h^2 \\
h^3+ih^4
\end{array} \right)
\quad {\rm and} \quad H^\dagger \tensor{D}^{\mu} H = H^\dagger D^\mu H -(D^\mu H)^\dagger H.
\end{equation}
Notice that the limiting case 
$g_\rho\approx 4\pi$ corresponds to a maximally strongly-coupled theory in the sense of naive 
dimensional analysis (NDA) \cite{Manohar:1983md}. However in the existing explicit models one understandably relies on a weaker coupling in order to retain computational control.
For example, in
the little Higgs theories \cite{Arkani-Hamed:2001nc} $g_\rho$ can be as weak as the typical SM coupling $g_\rho \sim g_{SM}$, while
in the holographic Higgs models \cite{Contino:2003ve} it is assumed $g_{SM} \alt g_\rho \alt 4\pi$. We should stress that even in
theories other than the composite Higgs models, it is still  useful to think of the new
physics in terms of the generic interaction strength $g_\rho$ and mass scales of new particles $m_\rho$. For instance in
supersymmetric theories, we have $g_\rho\sim g_{SM}$ and $f \sim  m_{SUSY}/g_\rho$.\footnote{However, in supersymmetry 
with R-parity and a fundamental Higgs scalar, the $c$'s in the first three lines of Eq.~(\ref{L_SILH}) are all one-loop suppressed $\sim g_{SM}^2/16\pi^2$. Similarly for KK-parity in extra-dimensional models.}

The coefficients $c_H$ through $c_g$ are expected to be of order 1 in a generic composite Higgs model. Counting of the power of $m_\rho$, $g_\rho$ and loop factors in Eq.~(\ref{L_SILH}) can then  be understood intuitively. 
For example, the terms in the first three lines are obtained from similar interactions of the SM by paying a factor of $1/f$ for each additional PNGB and a factor $1/m_\rho$
for each additional derivative. The terms in the fourth line are instead suppressed by a loop factor, which corresponds to assuming that in the limit $g_\rho < 4 \pi$ the underlying theory is an ordinary minimally coupled gauge theory, as is the case for little Higgs and holographic Higgs theories. Indeed the operators associated to $c_{HB}$ and $c_{HW}$ give rise to such
effects like an anomalous magnetic moment of the SM $W$ boson which in minimally coupled theories only arise at the loop level. Notice however that in the case of maximal NDA coupling $g_\rho \sim 4\pi$ the terms in the fourth line are not suppressed with respect to those in the third line. Finally the two operators appearing in the fifth line  both violate minimal coupling and break explicitly any 
Goldstone symmetry under which the Higgs boson shifts. This property explains our parametrization of their coefficient as we will better explain below. Notice that the PNGB nature of the Higgs was assumed only in the fifth line of Eq.~(\ref{L_SILH}).
 
 As discussed in detail in Ref.~\cite{Giudice:2007fh},
 the most relevant operators in Higgs physics  at the LHC are 
\begin{equation}
\label{pc:ope}
 \frac{c_H}{2f^2} {\cal O}_H, \quad  \frac{c_y y_f}{f^2} {\cal O}_y, \quad \frac{c_g g_s^2}{16\pi^2 f^2} \frac{y_t^2}{g_\rho^2}
{\cal O}_g, \quad \frac{c_\gamma g^{\prime\, 2}}{16\pi^2 f^2}\frac{g^2}{g_\rho^2} {\cal O}_\gamma .
\end{equation}
The operator associated with $c_T$ gives a contribution to 
to the $\rho$-parameter $\delta \rho = c_T v^2/f^2$, so we expect $c_T$ to be small either for symmetry reasons (e.g., custodial symmetry \cite{Chang:2003un, Chang:2003zn, Agashe:2004rs}, T-parity \cite{Cheng:2003ju,Cheng:2004yc,Low:2004xc}) or by simple tuning. The term $c_6$ in
Eq.~(\ref{L_SILH}) gives potentially important effects but only in Higgs self interactions, which are difficult to test at the LHC. 
The operators in the third line give subleading effects in the coupling to vector bosons, and moreover are already constrained by the LEP bounds on the $S$ parameter. The operators in the fourth line gives potentially sizable $O(v^2/f^2)$ effects, but only to observables like the decay rate of  $h\to Z\gamma$, which are  difficult to study at the LHC. 
Finally one should remark that for $g_\rho$ much bigger than the SM couplings, say $g_\rho \gg y_t$ like in the simple Georgi-Kaplan PNGB Higgs models \cite{Kaplan:1983fs,Kaplan:1983sm},  the effects of $c_g$ and $c_\gamma$ are suppressed relative to those of $c_H$ and $c_y$. But this is not the case in little Higgs models, in which $g_\rho \sim y_t$, and whenever there exists light colored or charged states with mass below $m_\rho$.

Let us analyze more closely the four relevant operators in Eq.~(\ref{pc:ope}).
The operator ${\cal O}_H$ contains four Higgs fields and two derivatives, and  is thus suppressed
by $1/f^2$ using the above power counting.  There are several possible sources for ${\cal O}_H$. First, integrating out heavy vectors and scalars coupling to the Higgs field at the tree-level will induce ${\cal O}_H$ regardless of whether the Higgs is a PNGB or not. Secondly, if the Higgs is
a PNGB, there are additional contributions from the nl$\sigma$m. 
In Section \ref{sect:OH} we will discuss these effects in detail.
On the other
hand, ${\cal O}_y$ breaks the same global symmetry as the SM Yukawa coupling, and has two more Higgs fields than the Yukawa 
coupling, which explain the $y_t/f^2$ suppression. The coefficient of ${\cal O}_g$ and ${\cal O}_\gamma$ can be understood in various ways. For instance the coefficient of  ${\cal O}_g$  can be written as
\begin{equation}
\frac{g_s^2}{m_\rho^2}\times \frac{g_\rho^2}{16\pi^2}\times \frac{y_t^2}{g_\rho^2}\,
\label{countcg}
\end{equation}
by which  $g_s^2$ counts the external SM gluon legs,  $1/m_\rho^2$ counts the operator dimension, $g_\rho^2/16\pi^2$ counts the fact that these operators arise at one-loop level in minimally coupled theories, and finally $g_{SM}^2/g_{\rho}^2$  represents the Goldstone boson suppression. In fact ${\cal O}_g$ and ${\cal O}_\gamma$ break the Goldstone shift symmetry $H_\alpha \to H_\alpha +c_\alpha$ in precisely the same way the Higgs mass term does. Eq.~(\ref{countcg}) then follows by assuming that no couplings,
 other than $g_{SM}$, break the Goldstone shift symmetry that protects the Higgs mass.  Therefore, quite generally, by symmetry and dimensional analysis
the coefficient of ${\cal O}_g$ can be estimated by attaching four derivatives (the field-strength squared) to the generic contribution to the Higgs mass term:
\begin{equation}
\delta m_H^2\, H^\dagger H \times \frac{g_s^2 }{m_\rho^4}G_{\mu\nu}G^{\mu\nu}\, .
\end{equation}
Assuming the heavy particle with mass $m_\rho$ cuts off the Higgs quadratic divergence, $\delta  m_H^2\sim g_{SM}^2m_\rho^2/16\pi^2$, we obtain again Eq.~(\ref{countcg}). Our power counting thus applies
to more general scenarios than composite Higgs models, such as  weak-scale supersymmetry.

It is worth commenting that in an unnatural theory effects of ${\cal O}_g$ and ${\cal O}_\gamma$ could be larger. In the case
 where the Higgs mass is light simply because of a tuning between the bare mass term and the
radiative corrections, $\delta m_H^2\sim g_{\rho}^2m_\rho^2/16\pi^2$, then the proper parametrization for ${\cal O}_g$ and ${\cal O}_\gamma$, barring additional tunings,  would be
\begin{equation}
\frac{\tilde c_\gamma g^{\prime\, 2}}{16\pi^2 f^2} H^\dagger H B_{\mu\nu} B^{\mu\nu},
  \quad \frac{\tilde c_g g_s^2}{16\pi^2 f^2}  H^\dagger H G_{\mu\nu} G^{\mu\nu} ,
  \end{equation}
  with $\tilde c_\gamma\sim \tilde c_g \sim 1 $.
The above result  gives sizable ${\cal O}(1)$ effects even when the lightest new states are rather heavy with $m_\rho \sim 4\pi f\sim 4\pi v \sim 2$ TeV.
Keeping this comment on unnatural theories in mind, in the rest of this section we will only focus on the PNGB case.

Aside from the above discussion of the physics underlying the SILH Lagrangian, a final technical comment is in order. The form of a Lagrangian can be changed, without affecting the physics, by reparametrizing the field variables. This well-known fact is particularly
important when dealing with effective Lagrangians based on a large mass expansion. In our case, reparametrizing the fields by treating $1/f^2$ as a small parameter, or, equivalently using the leading order equations of motion, the form of the dimension-six effective Lagrangian can be modified without affecting physics at ${\cal O}(1/f^2)$. In Eq.~(\ref{L_SILH}) a specific field parametrization was chosen so as to remove that redundancy. More specifically, by using the reparametrization freedom
\begin{equation}
\label{repara}
H\to H + a (H^\dagger H) H/f^2 ,
\end{equation}
the operator ${\cal O}_r =  H^\dagger H (D_\mu H)^\dagger (D^\mu H)$  was removed in the SILH Lagrangian \cite{Giudice:2007fh}.
Indeed it is easy to see that,  under
Eq.~(\ref{repara}), the Higgs kinetic term transforms as
\begin{equation}
(D_\mu H)^\dagger (D^\mu H) \to (D_\mu H)^\dagger (D^\mu H) + \frac{a}{f^2}\partial^\mu (H^\dagger H)\partial_\mu(H^\dagger H)
 + \frac{2a}{f^2} H^\dagger H (D_\mu H)^\dagger (D^\mu H) .
\end{equation}
while the Yukawa interaction transforms as 
\begin{equation}
- y_f \bar f_LHf_R \to - y_f\left(1+a\frac{H^\dagger H}{f^2}\right) \bar f_L H f_R\, .
\end{equation}
We thus deduce the useful operator identity (satisfied on the SM equations of motion, and neglecting the Higgs potential)
\begin{equation}
2{\cal O}_r={\cal O}_y-{\cal O}_H.
\label{ryH}
\end{equation}
Equivalently defining the coefficient of ${\cal O}_r$ to be $c_r/(2f^2)$, then under reparametrization of the Higgs field 
in Eq.~(\ref{repara}) we have
\begin{equation}
\label{under_a}
c_H \to c_H +2a , \quad c_r \to c_r +4a , \quad c_y \to c_y -a ,
\end{equation}
with all other coefficients unaffected.
Notice that the top Yukawa coupling in the SM is defined as 
$-y_t \bar{f}_L H f_R$ in our convention, hence the sign of the shift in $c_y$.
Starting from a generic Lagrangian with $c_r\not =0$, the SILH basis is obtained by choosing $a=-c_r/4$. Therefore parameters in the
SILH basis are related to those in a general basis by 
\begin{equation}
\left. \phantom{C^{A^a}_{B_B}}c_H\right\vert_{SILH}=c_H-\frac{c_r}{2}, \qquad\qquad \left. \phantom{C^{A^a}_{B_B}}c_y\right\vert_{SILH}=c_y+\frac{c_r}{4}.
\end{equation}
Physical amplitudes should depend only on reparametrization-invariant combination of parameters.

Eq.~(\ref{repara}) is an instance of a reparametrization of the dynamical variables in the composite sector to which the Higgs is assumed to belong. On the other hand, the vectors and fermions of the SM are assumed to be elementary. Technically this means (assumption 2 at the beginning of this section) that there exists a parametrization of the corresponding fields such that, neglecting the small effects of Yukawa couplings, all new physics effects are described by ``oblique'' operators which involve only
vector bosons and Higgses.
Eq.~(\ref{L_SILH}) is derived by working in such a basis, which is the same basis as employed in Ref.~\cite{Barbieri:2004qk}. One advantage of working in this basis is that there are no corrections to vertices
between SM vectors and fermions, allowing for a simple parametrization of the electroweak precision parameters. A generic operator basis can be obtained from ours by performing the following covariant redefinition of low-energy gauge fields:
\begin{eqnarray}
\label{eq:lowgauge}
g'B_\mu&\to& g'B_\mu +\frac{a_B}{f^2} H^\dagger \tensor{D}_{\mu} H , \\
\label{eq:lowgauge1}
gW^a_\mu&\to& gW^a_\mu+\frac{a_W}{f^2}H^\dagger \sigma^a \tensor{D}_\mu H ,
\end{eqnarray}
under which $c_H$, $c_r$, and $c_T$ shift. Moreover,  
from the gauge-covariant derivative in the fermion kinetic term, operators such as
 $H^\dagger \tensor{D}_{\mu} H (\bar{f}_L \gamma^\mu f_L)$ are generated, which
  modify SM vector-fermion vertices. When integrating out heavy
  vectors by working in the mass eigenbasis such terms are generically obtained. On the other hand, by working in the interaction eigenbasis \cite{Barbieri:2004qk, Marandella:2005wd}, that is defining the low-energy vector as the one that couples to the SM fermion, only oblique operators are generated. Needless to say, the low-energy physics does not depend on which 
interpolating fields we choose, as long as the one we keep fixed (as the low-energy field) has a non-vanishing matrix
element $\langle W_{SM}| W_{int} | 0\rangle$ between the vacuum and the light gauge boson.

Using the parametrization in Eq.~(\ref{L_SILH}) we can compute on-shell amplitudes of production, decay, and scattering processes. For example, the Higgs decay to fermions can be shown to be\footnote{We use the notation $H$ for the $SU(2)_L$ doublet scalar and $h$ for the neutral
component of $H$ which gets a VEV: $h\to h+v$.}
\begin{equation}
\label{amp:hff}
\Gamma(h \to f\bar{f}) \propto \left(\frac{y_f}{\sqrt{2}}\right)^2 \left[ 1 - \xi ( c_H+ 3c_y) \right],
\end{equation}
where $\xi \equiv v^2/f^2$.\footnote{More generally, in a basis where $c_r$ is non-vanishing, the coefficient of $\xi$ in 
Eq.~(\ref{amp:hff}) is replaced
by $ c_H + c_r/4 +3 c_y$, which is easily seen to be reparametrization invariant under Eq.~(\ref{under_a}).} However, we are more 
interested
in  comparing the results  to the corresponding SM predictions, which requires absorbing the ${\cal O}(v^2/f^2)$ contributions to the SM input
parameters such as the fermion masses $m_f$, Higgs mass $m_h$, and Fermi decay constant $G_F=1/v^2=(246\  {\rm GeV})^{-2}$. Below we quote from Ref.~\cite{Giudice:2007fh}
three amplitudes that most concern us in this study:
\begin{eqnarray}
\Gamma(h\to f\bar{f}) &=& \Gamma(h\to f\bar{f})_{SM}[ 1-\xi(c_H+2c_y)], \\
\label{eq:onshellhgg}
\Gamma(h\to gg) &=& \Gamma(h\to gg)_{SM} \left[ 1-\xi\, {\rm Re}\left( c_H + 2c_y + \frac{4y_t^2 c_g}{g_\rho^2 I_g} \right) \right] , \\
\Gamma(h\to \gamma\gamma) &=& \Gamma(h\to \gamma\gamma)_{SM} 
    \left[ 1- \xi \, {\rm Re} \left(\frac{c_H+2c_y}{1+J_\gamma/I_\gamma}\right.\right. \nonumber \\
  && \left.\left. \quad \quad \quad\quad \quad \quad
 + \frac{c_H-(g^2/g_\rho^2) {c}_W}{1+ I_\gamma/J_\gamma} + \frac{(4g^2/g_\rho^2)c_\gamma}{I_\gamma+J_\gamma}
  \right)\right], 
\end{eqnarray}
where the loop functions $I$ and $J$ are compiled in Appendix C of Ref.~\cite{Giudice:2007fh}. The above relations make clear
why we focus on  the four operators in Eq.~(\ref{allope}): these are the four operators controlling
the partial decay widths of the Higgs to $f\bar{f}$, $gg$, and $\gamma\gamma$. In $\Gamma(h\to \gamma\gamma)$ there is also 
a contribution of ${c}_W $, which can compete with those from $c_H$ and $c_y$ for $g_\rho\sim g_{SM}$. 

The power counting in Eq.~(\ref{pc:ope}) helps us understand the relative importance of various operators in the
on-shell amplitudes. For example, in composite Higgs theories where $g_\rho \agt g_{SM}$, the largest effect in
the on-shell amplitudes would come from ${\cal O}_H$ and ${\cal O}_y$, which are sensitive to the strongly-interacting sector 
responsible for breaking the electroweak symmetry \cite{Giudice:2007fh}. On the other hand, ${\cal O}_g$
and ${\cal O}_\gamma$ can potentially compete with ${\cal O}_H$ and ${\cal O}_y$ when there are light states actively involved in
the cancellation of the quadratic divergence in $m_H^2$.  In our parametrization this corresponds to the situation in which $g_\rho \sim g_{SM}$ and $m_\rho \sim g_{SM} f$.

\section{Theoretical Constraints on ${\cal O}_H$}
\label{sect:OH}

The operator ${\cal O}_H$ has two main phenomenological consequences \cite{Giudice:2007fh}. 
First, after $H$ gets a VEV, it gives an additional contribution to the neutral Higgs
kinetic term. Canonically normalizing the kinetic term amounts to a
rescaling of the neutral Higgs field
 by the factor $1/\sqrt{1+\xi c_H}\simeq 1-\xi c_H/2$, which reduces the single Higgs on-shell couplings by the same factor
 relative to the SM expectation. Secondly ${\cal O}_H$ modifies the high-energy behavior of the scattering amplitudes involving longitudinal vector bosons (and neutral Higgses). In unitary gauge this follows 
from the modification of the $hVV$ coupling, implying Higgs exchange does not perfectly unitarize scattering amplitudes \cite{Giudice:2007fh} . It is also evident working with the Goldstone bosons via the equivalence theorem \cite{Chanowitz:1985hj}: ${\cal O}_H$ implies scattering amplitudes growing like $E^2$.  Indeed its coefficient $g_\rho^2/m_\rho^2$ shows that ${\cal O}_H$ directly tests the non-linearity of the Higgs sector.

In this section we study a positivity constraint on $c_H$, and its exceptions.
We consider three contributions to $c_H$: 
1) from a nl$\sigma$m, 2) from integrating out a heavy
scalar, and 3) from integrating out a heavy vector. It can be shown, using group-theoretic methods that, 
contributions to $c_H$ from 1) and 3) are always positive. We will also consider unitarity constraints
on the sign of $c_H$ using dispersion relations, which reveal 
the contribution from 2) is positive except when there is a doubly-charged scalar such as an $SU(2)_L$ triplet.
Statements 2) and 3)  are independent of the composite nature of the Higgs boson and apply to models
with a fundamental scalar as well.

\subsection{Contributions From Non-linear Sigma Models}
\label{subsec:nlsm}

The most general nl$\sigma$m Lagrangian based on a coset $G/H$, where $G$ is the full symmetry group and $H$ the unbroken subgroup, can be written
down using the CCWZ formalism \cite{Coleman:sm,Callan:sn}. We will use the notation $T^i$ for the unbroken group generators in $H$ and $X^a$ the
broken generators in $G/H$. Since there is no gauge-covariant derivative in ${\cal O}_H$, we will turn off all the gauge fields in this
subsection. The PNGB is parametrized by the matrix
\begin{equation}
\label{eq:nonli}
 U= e^{i {\Pi}/f}, \quad \Pi=  \Pi^a X^a .
\end{equation}
The Goldstone-covariant derivative and the associated gauge field are derived from the Cartan-Maurer one-form \cite{Coleman:sm,Callan:sn, Weinberg:1996kr}:
\begin{eqnarray}
U^\dagger \partial_\mu U &=& i\frac1f \partial_\mu \Pi + \frac1{2f^2}[\Pi, \partial_\mu\Pi]-\frac{i}{6f^3}
[\Pi,[\Pi, \partial_\mu \Pi]] +{\cal O}(\Pi^4) \nonumber\\
&=& i {\cal D}_\mu ^a X^a + i {\cal E}_\mu^i T^i \equiv i {\cal D}_\mu + i {\cal E}_\mu .
\end{eqnarray}
We will consider a general coset space that is not necessarily symmetric, where the generators satisfy the following  commutation relations:
\begin{equation}
\label{eq:liealg}
[T^i, T^j]=if^{ijk} T^k, \quad [T^i, X^a]=i f^{iab} X^b ,\quad [X^a, X^b]=i f^{abi} T^i +i f^{abc} X^c.
\end{equation}
The coset space used in majority of the composite Higgs models is symmetric, for which $f^{abc}=0$ in the above,
except for the little Higgs model based on simple 
groups \cite{Kaplan:2003uc}.
Then it is straightforward to work out
\bea
\label{eq:Dungauge}
{\cal D}_\mu &=&\frac1{f} \partial_\mu \Pi   + \frac1{2f^2}[\Pi, \partial_\mu\Pi]_X-\frac{i}{6f^3}
[\Pi,[\Pi, \partial_\mu \Pi]]_X \nonumber \\
&=&X^a \left[\frac1{f} \partial_\mu \Pi^a +
 \frac{1}{2f^2} f^{abc}\Pi^b \partial_\mu \Pi^c  + \frac{1}{6f^3} 
(f^{cde} f^{bea} + f^{cdi} f^{bia})\Pi^b\Pi^c\partial_\mu \Pi^d\right],
\eea
where $[X^a, X^b]_{X,T}$ is the projection of the commutator in the broken and unbroken generators, respectively.
We have also dropped terms that are ${\cal O}(1/f^4)$ or higher. 
The leading two-derivative interaction term in the Lagrangian is
\begin{equation}
\label{lag}
\frac{f^2}2 \text{Tr} ({\cal D}^\mu {\cal D}_\mu) =  \frac12 \partial_\mu \Pi^a \partial^\mu \Pi^a 
 -\left(\frac1{6f^2} f^{aci}f^{bdi} + \frac1{24f^2} f^{ace}f^{bde}\right) \Pi^a\Pi^b\partial_\mu\Pi^c\partial^\mu\Pi^d .
\end{equation}
In the above we have chosen to normalize the group generators as 
${\rm Tr}(T^A T^B)={\rm Tr}(X^A X^B)=\delta^{AB}$. Notice that the only possible trilinear term 
$f^{abc} \partial_\mu \Pi^a \Pi^b \partial^\mu \Pi^c$ vanishes identically by Bose symmetry.

For our purpose, we will concentrate on $\Pi = h^a X^a$, where $X^a, a=1,2,3,4$, are the generators corresponding
to the Higgs field. Operators with two derivatives and four Higgses come from the last term in Eq.~(\ref{lag}):
\begin{eqnarray}
\label{eq1}
\frac{f^2}2 \text{Tr} ({\cal D}^\mu {\cal D}_\mu) &\supset& \frac1{f^2}h^a \ h^b\ \partial_\mu h^c\ \partial^\mu h^d \, {\cal T}^{abcd} ;  \\
\label{t1}
 {\cal T}^{abcd}&\equiv&-\left(\frac1{6f^2} f^{aci}f^{bdi} + \frac1{24f^2}f^{ace}f^{bde}\right).
\end{eqnarray}
The  four scalar fields $\{h^a, a=1,2,3,4\}$ together transform as a complex doublet under the electroweak group 
$SU(2)_L\times U(1)_Y$,  
and ${\cal T}^{abcd}$ is a fourth-order invariant tensor under that group. 
To see what operators arise from Eq.~(\ref{eq1}), it is most convenient to use 
the generators of the full custodial group $SO(4)\simeq SU(2)_L\times SU(2)_R$, bearing in mind that the custodial symmetry
is broken by the gauging of the hypercharge $Y=T_R^3$, where $T_L^A$ and $T_R^A$ are respectively the generators 
of $SU(2)_L$ and $SU(2)_R$ and $A=1,2,3$.
The Higgs field $\vec{h}=(h^1, h^2, h^3,h^4)^T$ transform like a vector as  $\mathbf{4}$ under the $SO(4)$. 
Thus ${h}^a\partial_\mu {h}^c$ in Eq.~(\ref{eq1}) contains the product
\begin{equation}
\label{eq:4x4}
\mathbf{4}\times \mathbf{4}= \mathbf{6}_A \oplus (\mathbf{1}\oplus\mathbf{9})_S ,
\end{equation}
where $S/A$ refers to the (anti-)symmetry property of the representation under the interchange of the two vectors. Under
$SU(2)_L\times SU(2)_R$, $\mathbf{6}_A$ is the adjoint representation $(\mathbf{3}_L, \mathbf{1}_R)\oplus(\mathbf{1}_L,
\mathbf{3}_R)$ and $\mathbf{9}_S=(\mathbf{3}_L, \mathbf{3}_R)$.
Since the structure constant is totally anti-symmetric,
the $(ac)$ and $(bd)$ components in ${\cal T}^{abcd}$ are also anti-symmetric and must live in the adjoint of $SO(4)$. We thus have
\begin{equation}
{\cal T}^{abcd}=\alpha_L (T_L^A)_{ac}(T_L^A)_{bd}+\alpha_R (T_R^A)_{ac}(T_R^A)_{bd} +\beta (T_R^3)_{ac}(T_R^3)_{bd}
\end{equation}
 The coefficient $\beta$,  associated with the hypercharge generator $T_R^3$, 
 vanishes for custodially invariant cosets. Using the explicit expression in the Appendix for the generators we have
\begin{eqnarray}
\label{eq:nlsmT}
{\cal T}^{abcd}&=&\alpha_+\left(\delta^{ab}\delta^{cd} - \delta^{ad}\delta^{bc}\right)+\alpha_-\epsilon_{abcd}-\frac{\beta}4 E_{ac}E_{bd}\label{Texplicit}\\
E_{ac}&=& (\delta^{a1} \delta^{c2} + \delta^{a3}\delta^{c4}) - (a \leftrightarrow c)\label{Eexplicit}
\end{eqnarray}
and  $\alpha_\pm=-(\alpha_R\pm \alpha_L)/4$. The parity odd term proportional to $\alpha_-$ vanishes by Bose symmetry when contracted in Eq.~(\ref{eq1}). The other terms contain a contribution to $c_r$ and, upon using the operator equivalence in Eq.~(\ref{ryH}), give the following combination of ${\cal O}_H$, ${\cal O}_y$ and ${\cal O}_T$
\begin{equation}
-{3}\frac{\alpha_+}{f^2} {\cal O}_H+{2}\frac{\alpha_+}{f^2} {\cal O}_y- \frac{\beta}{4f^2}{\cal O}_T
\end{equation}
By comparing Eqs.~(\ref{t1}), (\ref{Texplicit}), and (\ref{Eexplicit}) we deduce that 
$\alpha_+$ is negative. Indeed Eq.~(\ref{t1}) implies ${\cal T}^{1133}<0$ while Eqs.~(\ref{Texplicit}) and (\ref{Eexplicit})
give ${\cal T}^{1133}=\alpha_+$. Thus the nl$\sigma$m gives contributions
\beq
\label{eq:chnlsm}
 c_H^{(\sigma)}=-6\alpha_+>0, \quad c_y^{(\sigma)}=2\alpha_+<0, \quad
c_H^{(\sigma)}+2c_y^{(\sigma)} = -2\alpha_+ > 0.
\eeq 
Notice that the combination  $c_H^{(\sigma)}+2c_y^{(\sigma)}$, which controls the partial
widths $\Gamma(h\to \bar f f)$ and $\Gamma(h\to gg)$, is positive.

\subsection{Dispersion Relations}
\label{subsect:dis}
The sharp  result on the positivity of $c_H^{(\sigma)}$ that we just established naturally leads us to wonder whether  it is the general principles of quantum field theory that constrain the sign of $c_H$, irrespective of the PNGB nature of the Higgs.
 The usual way to investigate this question is to use dispersion relations in conjunction with reasonable hypotheses for the high energy  behavior of amplitudes and cross sections. We will do so in this section. The result, expressed in Eq.~(\ref{Dispersion relation}), is that  $c_H$ can  in principle become negative,  when
the total cross section among same charge Goldstones $\sigma_{++}$ exceeds the cross section among opposite sign
Goldstones $\sigma_{+-}$ over a significant range of energy.
The positivity bound in the previous subsection suggests this cannot be the case in an exact nl$\sigma$m. 
Negative contributions to $c_H$ could be possible only at an energy scale where the global symmetry is badly broken
and the nl$\sigma$m structure is completely lost. 
In the class of theories described at low-energy by our effective Lagrangian, this situation is realized at the scale $m_\rho \sim g_\rho f$ when the symmetry breaking splittings
$\sim g_{SM} f$ are comparable to $m_\rho$ itself, that is for $g_\rho \sim g_{SM}$. This situation is not realized in either the 
Georgi-Kaplan type or holographic Higgs models, but it can be realized in principle in little Higgs models.
In the following section we will indeed confirm that negative contributions to $c_H$ arise in such theories upon integrating out the scalar triplets. This nicely fits with the finding from dispersion relations, because the triplet 
contains a doubly charged scalar $\Phi_{++}$ which enhances  $\sigma_{++}$. 
On the other hand, even if such negative contributions exist, we will argue in Section \ref{sect:discussion} that
the overall contributions to $c_H$ should still be positive in little Higgs theories.

In order to proceed with dispersion relations it is instructive to first write down the low-energy  $a+b\to c+d$ scattering amplitude implied by Eq.~(\ref{eq1}). Taking with all momenta incoming, $p_a+p_b+p_c+p_d=0$, we find
\begin{eqnarray}
\	&&\left<a,\, b\right|  \frac1{f^2}h^\alpha \ h^\beta\ \partial_\mu h^\gamma\ \partial^\mu h^\delta \, {\cal T}^{\alpha\beta\gamma\delta}   \left|c,\, d \right>\nonumber\\
\	&& =\frac{s-u}{f^2}\left( {\cal T}^{abcd}+{\cal T}^{badc}\right)+
  \frac{s-t}{f^2}\left( {\cal T}^{abdc}+ {\cal T}^{bacd}\right) +\frac{t-u}{f^2}\left( {\cal T}^{acbd}+ {\cal T}^{cadb}\right)\,\,, 
\end{eqnarray}
where the Mandelstam variables are
\begin{eqnarray}
\	s=2p_a\cdot p_b=2p_c\cdot p_d, \qquad t=2p_a\cdot p_c=2p_b\cdot p_d,  \qquad u=2p_a\cdot p_d=2p_b\cdot p_c\,\,,
\end{eqnarray}
and we have used the antisymmetric properties of ${\cal T}^{abcd}$ encoded in Eq.~(\ref{t1}). The forward amplitude $(t=0)$ then reads
\begin{equation}
\	{\cal A}_{ab\to cd}(t=0)=\frac{s}{f^2}\left(2{\cal T}^{abcd}+{\cal T}^{abdc}+{\cal T}^{bacd}+2{\cal T}^{badc}+{\cal T}^{acbd}+{\cal T}^{cadb} \right) \,\,,
\end{equation}
so that one finds 
\beq
{\cal A}_{ab\rightarrow ab}(t=0)=0 ,\quad  {\cal A}_{ab\rightarrow ba}(t=0)=\frac{6s}{f^2}\,{\cal T}^{aabb} ,\quad
 {\cal A}_{aa\rightarrow bb}(t=0)=-\frac{6s}{f^2}\,{\cal T}^{aabb} .
\eeq
From the previous subsection we have $ {\cal A}_{13\rightarrow 31}(t=0)=-{\cal A}_{11\rightarrow 33}(t=0)=4c_H s/f^2$, so that the first derivative of these two amplitudes at $s=0$ (in particular the sign) is directly related to $c_H$.
Unfortunately these amplitudes are not fully elastic, because of the change of flavor quantum numbers between initial and final states. The elastic amplitude ${\cal A}_{ab\rightarrow ab}(t=0)$ vanishes exactly and thus carries no information on $c_H$. A non-vanishing forward amplitude can however be obtained by considering 
the scattering of combination of definite electrical charge. For these amplitudes it is not possible to disentangle the contribution of $c_H$ and $c_T$, and we will therefore focus on the phenomenologically relevant case of $c_T=0$.
Using the notation $\pi^{\pm}=(h_1\pm ih_2)/\sqrt 2$, $h_4=\pi^0$, and $h_3=h$ we have
\beq
-{\cal A}\left( \pi^0 \pi^0 \to \pi^+ \pi^- \right) =
-{\cal A}\left( \pi^+ \pi^- \to \pi^0 \pi^0 \right)=
{\cal A}\left( \pi^\pm \pi^\pm \to \pi^\pm \pi^\pm \right)=-\frac{c_H s}{f^2},
\label{strsc}
\eeq
\beq
{\cal A}\left(  \pi^\pm \pi^0 \to  \pi^\pm \pi^0\right)=\frac{c_H t}{f^2},~~~~
{\cal A}\left( \pi^+  \pi^- \to  \pi^+  \pi^- \right) =-\frac{c_H u}{f^2},
\label{strsc2}
\eeq
and the same amplitudes upon exchanging  $\pi^0$ with $ h$. The elastic amplitudes, ${\cal A}\left( \pi^+ \pi^+ \to \pi^+ \pi^+ \right)$ and $ {\cal A}\left( \pi^+  \pi^- \to  \pi^+  \pi^- \right)$, are related by crossing symmetry $s \leftrightarrow u$ and do not vanish at $t=0$. Using them we can thus derive a potentially useful dispersion relation. 
Defining  ${\cal A}_{++}(s)\equiv {\cal A}\left( \pi^+ \pi^+ \to \pi^+ \pi^+ \right)(s,t=0)$ and ${\cal A}_{+-}(s)\equiv {\cal A}\left( \pi^+  \pi^- \to  \pi^+  \pi^- \right)(s,t=0)$, we will assume 
\begin{equation}
\displaystyle\lim_{s\rightarrow\infty} \frac{\mathcal{A_{+-}}(s)}{s}={\rm constant}\equiv c_\infty \ .
\end{equation}
This assumption allows us to include the case where charge neutral massive  vector fields are exchanged in the $t$-channel. In that case by direct computation one finds $c_\infty > 0$. We can thus derive a subtracted dispersion relation
\begin{eqnarray}
\ \mathcal{A'_{+-}}(s)-c_\infty &&=\frac{1}{2\pi i}\oint \frac{\mathcal{A_{+-}}(z)-c_\infty z}{(z-s)^2}dz= \frac{1}{2\pi i}\left(\int_{-\infty}^0+\int_{4m^2}^\infty\right) \frac{\mbox{Disc}(\mathcal{A_{+-}}(x))}{(x-s)^2}dx\nonumber\\
\ && = \frac{1}{2\pi i}\int_{4m^2}^\infty \left(\frac{\mbox{Disc}(\mathcal{A_{+-}}(x))}{(x-s)^2}+\frac{\mbox{Disc}(\mathcal{A_{+-}}(4m^2-x))}{(x-4m^2+s)^2}\right)dx
\label{dispersion}
\end{eqnarray}
where we have assumed $\pi^\pm$ has a {\it small} mass $m^2$, which will be sent to zero later.
In the last equality, in order to reduce to a single integral, we have performed the change of variable $x\to 4m^2-x$. 
By crossing symmetry we have
\begin{equation}
{\cal A}_{+-}(4m^2-s)={\cal A}_{+-}(u)={\cal A}_{++}(s) ,
\end{equation}
and therefore
\begin{eqnarray}
\mbox{Disc}({\cal A}_{+-}(4m^2-s))&=&{\cal A}_{+-}(4m^2-s+i\epsilon)-{\cal A}_{+-}(4m^2-s-i\epsilon)\nonumber \\
&=&{\cal A}_{++}(s-i\epsilon)-{\cal A}_{++}(s+i\epsilon)
= -\mbox{Disc}({\cal A}_{++}(s)) .
\end{eqnarray}
Applying the optical theorem we thus have
\bea
 \mbox{Disc}(\mathcal{A_{+-}})(s)\big |_{s>0}&=& \phantom{-}2i\,\mbox{Im}(\mathcal{A_{+-}})(s)= \phantom{-}2i\sqrt{s(s-4m^2)}\ \sigma_{+-} , \\
  \mbox{Disc}(\mathcal{A_{+-}})(s)\big |_{s<0}&=& -2i\,\mbox{Im}(\mathcal{A_{++}})(-s)=-2i\sqrt{s(s-4m^2)}\ \sigma_{++} ,
  \eea
where $\sigma_{ij}$ refers to the total cross section for the process $ij \longrightarrow \mbox{everything}$. Notice the sign
flip in the crossed channel $++\to ++$. Thus by
 considering Eqs.~(\ref{strsc2}) and (\ref{dispersion}) at $s=0$ and taking the small mass $m=0$ we can write
 \begin{eqnarray}\label{Dispersion relation}
\ c_H =c_\infty +\frac{f^2}{\pi } \int_{0}^\infty \left(\sigma_{+-}(x)-\sigma_{++}(x)\right)\frac{dx}{x}
\end{eqnarray}
While this result shows that $c_H$ is not  positive in the most general case, it is still quite useful, as it constrains the sources of  negative contributions.
These must either come from the far UV, via $c_\infty$, or via a sizeable cross section in the $++$ channel. We have already mentioned that when vector boson exchange dominates the UV, one has  $c_\infty>0$. In that situation, negative contributions can only be due to $\sigma_{++}$. At tree level such a contribution
can  only come by integrating out an object with charge $+2$. Since there exists no coupling of a vector with two equally charged objects, we conclude that only by  integrating out a scalar multiplet containing a charge $+2$ state can one get a negative contribution to $c_H$ at  tree level. This is what we willl prove explicitly in the next section.

\subsection{Contributions From Heavy Scalars}
\label{sect:scalarcont}
 
In this subsection we consider effects on $c_H$ and $c_T$, as well as the associated contribution to $c_y$,  
from integrating out heavy scalars at tree level. The treatment here on does not make use of the composite nature of the Higgs, 
except when we discuss the specific contribution from little Higgs theories toward the end of this subsection.

Since we are interested in dimension-six operators with two-derivative and four-Higgs that are induced from integrating out a heavy
scalar at the tree evel, we only need to consider cubic interactions involving two Higgses and one heavy scalar. At leading order these trilinears are associated with the scalar potential, since the nl$\sigma$m 
structure implies no trilinears involving two derivatives, as commented below Eq.~(\ref{lag}) in
 Section \ref{subsec:nlsm}.

Cubic interactions involving two Higgses can be classified according to the transformation properties under 
$SU(2)_L\times U(1)_Y$.  
It will be convenient to again use the $SO(4)$ notation introduced in Section \ref{subsec:nlsm}.
In this case the symmetric product $\vec{h}\vec{h}^T$ can be decomposed either in 
one real singlet $\phi_s$, one real triplet 
$\phi^{a}_r$, or one complex triplet $\phi^{a}_c$:
\begin{eqnarray}
\label{eq:tripletform}
&& \phi_s \sim \vec{h} \cdot \vec{h} = H^\dagger H , \nonumber \\
&& \phi^{a}_r \sim \vec{h}^T \ T_L^a T_R^3\  \vec{h} = H^\dagger \frac{\sigma^a}2 H ,\\
&& \phi^{a}_c \sim  \vec{h}^T \ T_L^a (T_R^1- i T_R^2)\  \vec{h}=H^T \epsilon \frac{\sigma^a}2 H ,\nonumber
\end{eqnarray}
where $\epsilon = i \sigma^2$.
For custodially invariant theory $\phi^{a}_r$ and $\phi^{a}_c$ combine into a $(\mathbf{3}_L,\mathbf{3}_R)$:
$\phi^{AB} \sim \vec{h}^T \ T_L^A T_R^B\  \vec{h}$.
The invariance under the electroweak group dictates the heavy scalar field coupling to the Higgs through
the above cubic interactions can only be
a real singlet $\Phi_s$, a real triplet $\Phi_r^a$, or a complex triplet $\Phi^a_c$. The total contribution to the effective coupling $c_H$ and $c_T$ will be a sum from the three sectors without any interference.

Consider first  the case of a real triplet $\Phi^{a}_r$.
 The relevant part of the Lagrangian involving the triplet and the Higgs scalars is
\begin{equation}
\label{eq:realtriplet}
{\cal L}_s = -\frac12\Phi^{a}_r \Box \Phi^a_r -\frac12 m_r^2 \Phi_r^a \Phi_r^a +
 \beta_r f\, \Phi_r^a ( \vec{h}^T \ T_L^a T_R^3\  \vec{h}) +\cdots \ .
\end{equation}
 By integrating out $\Phi_r^a$ we find
\begin{eqnarray}
{\cal L}_{eff}&=&\frac{\beta_r^2f^2}{2}( \vec{h}^T \ T_L^a T_R^3\  \vec{h})\frac{1}{\Box +m_r^2}( \vec{h}^T \ T_L^a T_R^3\  \vec{h})\nonumber \\
&=&\frac{\beta_r^2f^2}{2m_r^2}( \vec{h}^T \ T_L^a T_R^3\  \vec{h})\left [1-\frac{\Box}{m_r^2}+\cdots\right ]( \vec{h}^T \ T_L^a T_R^3\  \vec{h}),
\end{eqnarray}
 where the first term in the expansion gives rise to the celebrated quartic coupling for the Higgs in little Higgs models,
 while the second, upon use of operator identity in Eq.~(\ref{ryH}) to remove ${\cal O}_r$, can be written as 
\begin{equation}
- \frac{\beta_r^2 f^2}{ 4 m_r^4}\left({\cal O}_H-{\cal O}_y-\frac12{\cal O}_T\right) .
\label{triplet0}
\end{equation}
Very much as we have treated the case of a neutral triplet we can treat the case of a complex triplet and a real singlet similarly.

For the complex triplet we start from
\begin{equation}
\label{eq:complextriplet}
{\cal L}_s = -\Phi^{a\, *}_c \Box \Phi^a_c - m_c^2 \Phi_c^{a\,*} \Phi_c^{a} +
 \beta_c f\, \Phi_c^{a\,*} ( \vec{h}^T \ T_L^a (T_R^1-iT_R^2)\  \vec{h}) + {\rm h.c.}\quad ,
\end{equation}
and by integrating out $\Phi^a_c$ we find, in addition to a contribution to the Higgs quartic coupling, 
\begin{equation}
 - \frac{ \beta_c^2 f^2}{ 4 m_c^4}\left ({\cal O}_H-2{\cal O}_y +{\cal O}_T\right ).
\label{triplet+}
\end{equation}
For the scalar singlet, the Lagrangian is
\begin{equation}
\label{eq:singlet}
{\cal L}_s = -\frac12\Phi_s \Box \Phi_s -\frac12 m_s^2 \Phi_s \Phi_s  +\beta_s f\, \Phi_s ( \vec{h}\cdot \vec{h})   ,
\end{equation}
which at low energy contributes to the following dimension six operator  
\begin{equation}
 + \frac{ \beta_s^2f^2}{ 2 m_s^4}{\cal O}_H .
\label{singlet0}
\end{equation}
Notice that in the case of a singlet, we could in principle have also added a tadpole term. In little Higgs models, as recently emphasized in Ref.~\cite{Schmaltz:2008vd}, if this tadpole has the size expected by symmetry considerations, then 
the very mechanism of suppression of the Higgs mass fails. 
We thus consider the case of a singlet just for completeness.

From the above we find that $c_H$ is always positive
for a singlet scalar and negative for a triplet scalar, while for $c_y$ is positive for triplets and vanishing for singlets.
Summing all scalar contributions in Eqs.~(\ref{triplet0}), (\ref{triplet+}),
and (\ref{singlet0}) we find 
\bea
\label{totalscalar}
c_H^{(s)} &=& -\frac{\beta_r^2 f^4}{2m_r^4}  -\frac{\beta_c^2 f^4}{2m_c^4} + \frac{\beta_s^2 f^4}{m_s^4} ,\\
c_y^{(s)} &=& +\frac{\beta_r^2 f^4}{4m_r^4}+ \frac{\beta_c^2 f^4}{2m_c^4} >0 , \\
c_T^{(s)} &=& +\frac{\beta_r^2 f^4}{4m_r^4}-\frac{\beta_c^2 f^4}{2m_c^4}  , \\
\label{eq:scalarchcy}
c_H^{(s)}+2 c_y^{(s)} &=& + \frac{\beta_c^2 f^4}{2m_c^4}  + \frac{\beta_s^2 f^4}{m_s^4}>0  .
\eea
In a custodially invariant theory the neutral and complex triplets will combine together
into a $(\mathbf{3}_L,\mathbf{3}_R)$ representation of $SU(2)_L\times SU(2)_R$ and the parameters will satisfy the relation: $2\beta_c^2=\beta_r^2$ and $m_r^2=m_c^2$. In this case we see explicitly that $c_T^{(s)}=0$. As for the combination $c_H+2c_y$ which
enters into the on-shell coupling of the Higgs with the fermion, the scalar contribution is again always positive for the 
complex triplet and the singlet scalars, while it vanishes for the real triplet scalar.
Notice that the  negative $c_H$ contribution for triplets is in accordance with the finding using dispersion relations in Section \ref{subsect:dis}.

 The above result seemingly disrupts the positivity of $c_H$, 
which was deduced in Section \ref{subsec:nlsm} for a general nl$\sigma$m, while $c_H+2c_y$ is still positive.
However it turns out that in all little Higgs models the sum of all contributions to $c_H$ still remains positive, in spite of the negative contribution from triplet scalars.
We will discuss the combination of all contributions in Section \ref{sect:discussion}. For the moment it
is worth pointing out the general form of the coefficient
in Eq.~(\ref{totalscalar}) in typical little Higgs models. 
For instance, let's consider the littlest Higgs model \cite{Arkani-Hamed:2002qy} as an illustration.
The potential for the triplet scalar $\phi_{ij}$ is
\begin{equation}
\label{eq:scalarp}
 V(\phi,\phi^*) = g_L^2 f^2 \left| \phi_{ij} + \frac{i}{2f}(H_i H_j+H_j H_i)\right|^2 
   + g_R^2 f^2 \left| \phi_{ij} - \frac{i}{2f}(H_i H_j+H_j H_i)\right|^2 ,
   \end{equation}
where the notation is such that $H = (H_1, H_2)^T$. Since $H^T \epsilon \sigma^3 H=-(H_1 H_2+H_2 H_1)$, we see
that $\phi_{12}=\phi_{21}=\Phi_c^3/\sqrt{2}$. Therefore
 \begin{equation}
m_c^2 = (g_L^2+g_R^2) f^2 , \quad \beta_c= \sqrt{2}(g_R^2-g_L^2),\quad c_H^{(s)}= - \frac{(g_R^2-g_L^2)^2}{(g_R^2+g_L^2)^2}.
\eeq
The crucial observation here is that $c_H^{(s)}$ is maximally negative in the limit
in which one coupling, say $g_R$, is much bigger than the other. Away from this limit, $c_H^{(s)}$ is monotonically increasing
until it reaches zero at $g_L=g_R$. The special form of the scalar potential in Eq.~(\ref{eq:scalarp}) is dictated by the 
collective breaking mechanism and thus generic in little Higgs theories. Hence the preceding observation holds for all
little Higgs models. In Section \ref{sect:discussion} we will make use of this observation to argue that the negative $c_H^{(s)}$
in little Higgs theories cannot overcome other positive contributions so that in the end the overall $c_H$ stays positive.
It is also worth observing that in the T-parity limit, $g_R=g_L$, $c_H^{(s)}=0$ as expected since a cubic coupling between one triplet (T-odd) and two Higgs doublets (T-even) are forbidden.

\subsection{Integrating Out Heavy Vectors}
\label{vectorsSHORT}

In this subsection we present a general proof on the positivity of  $c_H$ from integrating out heavy vectors at tree level,
which applies to both cases of a fundamental and a composite Higgs scalars. In fact, we prove a stronger result, that
$c_H$ contains no effects of the order $g_{SM}/g_\rho$ when integrating out heavy vectors. This statement is another element in our
argument in Section \ref{sect:discussion} for the positivity of $c_H$ in little Higgs theories,
despite the negative contribution from the triplet scalar. It is also possible to calculate effects of integrating out heavy
vectors within a nl$\sigma$m explicitly, which is presented in Appendix B.


Similar to the case
of heavy scalars, the quantum number of the heavy vector is determined by classifying the low-energy operators, in this case the
vector current, made out of the light fields. For the Higgs current, hermiticity requires such vector bilinears to 
be anti-symmetric in exchange of the
two Higgs fields involved, which then must live in the adjoint 
$\mathbf{6}_A=(\mathbf{3}_L,\mathbf{1}_R)\oplus (\mathbf{1}_L,\mathbf{3}_R)$ of $SO(4)$:
\begin{eqnarray}
(\mathbf{3}_L,\mathbf{1}_R)&:& J_{HL\,\mu}^a \equiv \frac12 i\, \vec{h}^T \,T_L^a \tensor{D}_\mu \, \vec{h}=
    i H^\dagger \, \frac{\sigma}{2}^a \tensor{D}_\mu \, H \\
 (\mathbf{1}_L,\mathbf{3}_R)&\supset& J_{HR\,\mu}^3 = i \, \vec{h}^T\, T_R^3  \tensor{D}_\mu\, \vec{h}= i H^\dagger  \, \tensor{D}_\mu \,H  \\
 (\mathbf{1}_L,\mathbf{3}_R)&\supset& J_{HR\,\mu}^- = i \, \vec{h}^T\, (T_R^1-iT_R^2)  \tensor{D}_\mu\, \vec{h}= i \, H^T\,\epsilon  \tensor{D}_\mu \,H,
\end{eqnarray}
where $\vec{h}^T \,T_L^a \tensor{D}_\mu\,  \vec{h}\equiv\vec{h}^T\, T_L^a (D_\mu  \vec{h}) - (D_\mu  \vec{h})^T \,T_L^A\, \vec{h}$, and
similarly for the $\tensor{D}_\mu$ operator in other bilinears. Furthermore, $D_\mu \vec{h}$ is the gauge-covariant derivative with respect
to the electroweak $SU(2)_L\times U(1)_Y$. The current $J_{HR\,\mu}^+$, which is not written out explicitly above, can be
obtained by taking the complex-conjugate of $J_{HR\,\mu}^-$.
Vector currents with the same quantum numbers as $J_{HL\,\mu}^a$ and $J_{HR\, \mu}^3$ can also be constructed from light vectors and light fermions, respectively,
\begin{eqnarray}
(\mathbf{3}_L,\mathbf{1}_R)&:& J_{F\,\mu}^{a} \equiv (D^\nu W_{\nu\mu})^a \qquad {\rm and} \qquad 
J_{\psi\,\mu}^{a}\equiv \bar{\psi}\,\gamma_\mu \frac{\sigma}{2}^a \,\psi ,\\
 (\mathbf{1}_L,\mathbf{3}_R)&\supset& J_{F\,\mu}^{Y}\equiv
 D^\nu B_{\nu\mu} \qquad \quad\,  \, {\rm and} \qquad J_{\psi\,\mu}^{Y}\equiv \bar{\psi}\,\gamma_\mu Y\, \psi , 
\end{eqnarray}
where by $J_{\psi\,\mu}^{a}$ and $J_{\psi\,\mu}^{Y}$ we indicate collectively the quark plus lepton contributions to the weak isospin and hypercharge currents respectively. Notice that since $T_R^{1,2}$ is not gauged we cannot write a current with the same quantum numbers of $J_{LR\,\mu}^\pm$ by just using the SM vector fields. On the other hand, using fermions 
  it is possible to construct additional currents, for instance a right-handed charged current or  baryon/lepton currents. 
  However in this work we would like to stay with the so-called universal models, where the SM fermions interact only via the $SU(2)_L\times U(1)_Y$ currents, and do not consider these possibilities.

In Section \ref{sect:powercount} we discussed the possibility of choosing a specific definition of low-energy gauge 
fields, Eqs.~(\ref{eq:lowgauge}) and (\ref{eq:lowgauge1}), so that the light fermions appear only via 
operators associated with the Yukawa coupling, and not via operators that renormalize the vertices with the SM $W$ and $Z$ bosons.
This is equivalent to using the SM equations of motion
  \begin{eqnarray}
 (D^\nu W_{\nu\mu})^a+g J_{HL\mu}^a+gJ^a_\mu&=&0\\
 D^\nu B_{\nu\mu} +g'J_{HR\mu}^3+g'J^Y_\mu&=&0
 \end{eqnarray}
 to express the fermionic current in terms operators involving only the Higgs and vector fields.
  Under the above assumptions,  the most general Lagrangian describing the interactions of heavy vectors with the SM fields, at lowest order in a large mass expansion, is written as
  \begin{eqnarray}
 {\cal L}_{V}&=&\frac{m_L^2}{2}V^{a\mu}V^a_\mu+ V^{a\mu}\left [\gamma_H g_\rho J_{HL\mu}^a+\gamma_V \frac{g}{g_\rho} (D^\nu W_{\nu\mu})^a\right ] \nonumber  \\
 &&+\frac{m_0^2}{2}V^{0\mu}V^0_\mu+ V^{0\mu}\left (\delta_H g_\rho J_{HR\mu}^3+\delta_V  \frac{g'}{g_\rho}D^\nu B_{\nu\mu}\right )\nonumber \\
 &&+{m_+^2}V^{+\mu}V^-_\mu+\frac{g_\rho}{\sqrt 2}\left ( V^{+\mu} J_{HR\mu}^-+{\rm h.c.}\right ) \ .
 \end{eqnarray}
 Notice that according to the SILH hypothesis we expect the coefficients    $\gamma_H, \,\gamma_V, \,\delta_H,\,\delta_V$ to be all of order unity. 
Also given our choice of normalization, custodial symmetry in the sector containing the Higgs and heavy vectors corresponds to
 $m_0^2=m_+^2$ and $\delta_H=1$. Upon integrating out the heavy vectors we find
  \begin{eqnarray}
 {\cal L}_{V}&=&-\frac{1}{2 m_L^2}\left (\gamma_H g_\rho J_{HL\mu}^a+\gamma_V \frac{g}{g_\rho} D^\nu W^a_{\nu\mu}\right )^2
 \nonumber \\
 &&-\frac{1}{2m_0^2}\left (\delta_H g_\rho J_{HR\mu}^3+\delta_V  \frac{g'}{g_\rho}D^\nu B_{\nu\mu}\right )^2 \nonumber \\
 &&-\frac{g_\rho^2}{2m_+^2}J_{HR\mu}^-J_{HR}^{+\mu}.
 \label{heavyVectors}
 \end{eqnarray}
 The presence of perfect squares in the above equations fixes unambiguously the sign of the contribution to the coefficients of several operators. In addition, if we go to the non-canonical basis where the gauge coupling is removed from the covariant derivative by rescaling the gauge field 
 $g A_\mu \to A_\mu$, then it is clear that the coefficient of two-derivative operators such as $J^{\mu\,a} J^a_\mu$ can not depend on the gauge couplings; only operators with more derivatives can have coefficients dependent on the gauge couplings. More explicitly, two-derivative operators only come from interactions in the Lagrangian containing one or no derivative:
 \begin{equation}
 v_{ab}\, A^{\mu\,a}A_\mu^b + A^{\mu\,a}J_\mu^a \quad ,
 \end{equation}
 which do not involve gauge couplings in the non-canonical basis. Thus neither coefficient of ${\cal O}_H$ and ${\cal O}_r$ can depend on the gauge coupling.

Using the explicit form of the Higgs currents and the operator identity in Eq.~(\ref{ryH})
we get, from Eq.~(\ref{heavyVectors}),
 \bea
 \label{eq:chinvec}
 c_H& =& \frac{3 g_\rho^2 \gamma_H^2 f^2}{4 m_L^2}+ \frac{3 g_\rho^2 f^2}{ m_+^2} >0 , \\
c_y &=& - \frac{ g_\rho^2 \gamma_H^2 f^2}{4 m_L^2 } - \frac{ g_\rho^2 f^2}{ m_+^2} <0,\\
c_T &=& \frac{ g_\rho^2\delta_H^2 f^2}{ m_0^2} - \frac{ g_\rho^2 f^2}{ m_+^2} , \\
\label{eq:vecchcy}
c_H+2c_y &=& \frac{ g_\rho^2 \gamma_H^2 f^2}{4 m_L^2}+ \frac{ g_\rho^2 f^2}{ m_+^2} >0.
\eea 
Arguments in the preceding paragraph demonstrate that the above coefficients are all independent of gauge couplings.
We also have positive contributions to $c_{2W}\propto \gamma_V^2$ and $c_{2B}\propto \delta_V^2$, where ${\cal O}_{2W}=  (D^\nu W^a_{\nu\mu})^2$ and similarly for ${\cal O}_{2B}$,
 as was already noticed in Ref.~\cite{Barbieri:2004qk} and proven from prime principles in Ref.~\cite{Cacciapaglia:2006pk}. On the other hand cross terms are simply ${\cal O}_{W}$ and ${\cal O}_B$ in Eq.~(\ref{L_SILH}), which renormalize $c_W\propto \gamma_H\gamma_V$ and $c_B\propto \delta_H\delta_V$ and do not have a definite sign in general. Since $c_W+c_B$ is related to the $S$ parameter \cite{Giudice:2007fh},
we see a positivity constraint of $S$ is on less solid grounds. However, under specific assumptions about the structure 
 of the heavy vector Lagrangian, for instance the minimal coupling hypothesis of Ref.~\cite{Giudice:2007fh}, positivity of $c_W$ and $c_B$ holds as well.

\section{\label{detail} Theoretical Constraints on ${\cal O}_g$, ${\cal O}_{y}$, and ${\cal O}_\gamma$}
\label{sect:Og}

In this section we discuss constraints on ${\cal O}_g$ and ${\cal O}_\gamma$ from naturalness principle. Along the way
we make a remark on the sign of ${\cal O}_{y}$ which applies to all composite Higgs models we investigated. The statements
on $c_g$ and $c_\gamma$ are valid even for a fundamental Higgs scalar.

The naturalness principle
has been a prominent guiding principle for physics beyond the SM
in the past few decades. 
The largest contributions to the Higgs quadratic
divergences in the SM come from the Higgs self-interactions, the gauge bosons,
and the top quark:
\begin{eqnarray}
\delta m_h^2(h) &\simeq& \frac{\lambda}{16\pi^2}\, \Lambda^2 \sim (50 {\rm\ GeV})^2, 
\nonumber \\
\delta m^2_h(g) &\simeq& \frac{g^2}{16\pi^2}\, \Lambda^2 \sim (70 {\rm\ GeV})^2,
\nonumber \\
\delta m^2_h(t) &\simeq& -\frac{3\lambda_t}{8\pi^2}\, \Lambda^2 \sim 
  -(200 {\rm\ GeV})^2, \nonumber
\end{eqnarray}
where the numbers are obtained for $\Lambda \sim 1$ TeV. These numbers suggest that,
in order to stabilize the Higgs mass at a few hundreds GeV, new particles at or below
TeV scale should be present to cancel these quadratic divergences. In supersymmetric theories
 the new particles responsible for the cancellations have
opposite spin-statistics to their SM partners, while new particles in composite Higgs models
have the same spin-statistics to the SM partners. In this section we focus on the case when new
particles have the same spin-statistics as their SM partners, and briefly comment on supersymmetric
theories in the end.

From the observation that top quarks have the
largest contribution to the quadratic divergence, it is generally believed that partners of the
SM top quark should not be heavier than 1 TeV, while partners of the SM Higgs and gauge bosons
could be as heavy as a few TeV without re-introducing fine-tunings in the Higgs mass. Furthermore,
it is plausible that partners of top quarks would also carry $SU(3)_c$ and hence be copiously produced at
the LHC. In other words, if naturalness principle is upheld in the electroweak sector of the SM,
there may very well be new colored states 
as light as several hundreds GeV, which would have a great chance of being observed
directly and/or indirectly at the LHC. Conversely, if such new colored particles are observed, it
will be of great importance to verify if they indeed participate in the cancellation of Higgs
quadratic divergences.

\begin{figure}[t]
\includegraphics[scale=0.90,angle=0]{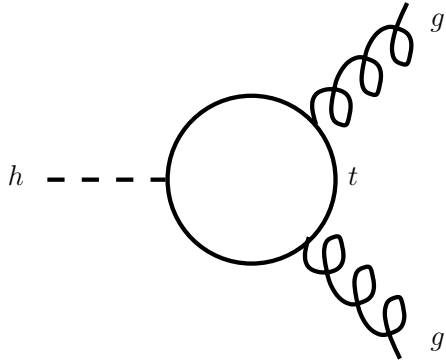}
\caption{\label{fig1}{\it Standard model top quark contribution to the gluon fusion production.}}
\end{figure}

In the SM it is well known that the top quark is the dominant contribution to the gluon fusion production
of the Higgs boson \cite{Gunion:1989we}, which is a loop induced process as shown in Fig.~\ref{fig1}. 
When the Higgs is light, $m_h^2/(4m_t^2) \alt 1$, the loop diagram can be approximated by a dimension-five
operator $(1/v)\ h G_a^{\mu\nu} G^{a\, \mu\nu}$ whose coefficient is related to the QCD beta
function through the low-energy theorems \cite{Ellis:1975ap,Shifman:1979eb}. Formally the 
effective theory has an expansion parameter in $m_h^2/(4m_t^2)$, however, actual computations of
QCD corrections suggest the approximation using an effective operator remains extremely good 
for a Higgs mass up to 1 TeV \cite{Spira:1995rr}. When there are new colored fermions with a significant
coupling to the Higgs boson\footnote{The new fermion could be vector-like and receive most of its mass
through a Dirac mass term, while still couple to the SM Higgs through Yukawa couplings.}, the
operator ${\cal O}_g$ will be induced at energies below the fermion 
mass scale, which gives a new contribution to $(1/v)\ h G_a^{\mu\nu} G^{a\, \mu\nu}$
 after electroweak symmetry breaking. In this section we will demonstrate
a correlation between the cancellation (or the lack thereof) of the Higgs quadratic divergences 
in the top sector and the sign of $c_g$ induced by the new heavy fermion. It will
also become clear that a similar analysis could be extended to $c_\gamma$ induced by a new set
of heavy electroweak gauge bosons.

\subsection{Higgs Low-Energy Theorems}
\label{sub:higgs}

We will now  flesh out the low-energy theorem which relates  the  Higgs-gluon vertex to the gluon two-point function arising from loops of heavy states. Throughout this subsection we work in unitary gauge with the Higgs doublet $H=(0,h/\sqrt 2)^T$. Upon electroweak symmetry breaking we will make the substitution $h\to h+v$ with $v\simeq 246 $ GeV. So the top Yukawa  is
\begin{equation}
\label{smtop}
{\cal L}_t = - \frac{\lambda_t}{\sqrt{2}}\, \bar{t}_L t_R h , \quad \frac{1}{\sqrt{2}}\lambda_t v = m_t.
\end{equation}
 Consider then a heavy colored multiplet whose mass $M$ depends on the Higgs expectation value $M\equiv M(h)$. In the limit
$m_h\ll M$ we can integrate out the heavy state and describe the coupling of $h$ to gluons via an effective Lagrangian  in a $1/M$ expansion. By gauge invariance, the leading interaction in the derivative expansion   has the form $f(h) G_{\mu\nu}^aG^{a\,\mu\nu}$. The function $f$ is fixed, 
up to an uninteresting scheme dependent additive constant, by matching the coupling constant in the low energy and high energy theory \cite{Weinberg:1980wa}
\begin{equation}
{\cal L}_{eff}= -\frac{1}{4}\frac{1}{g_{eff}^2(\mu,h)}G_{\mu\nu}^a G^{a\,\mu\nu}=-\frac{1}{4}\left (\frac{1}{g_s^2(\mu)}- b \frac{t_r}{4\pi^2}\log \frac{M(h)}{\mu}\right )G_{\mu\nu}^aG^{a\,\mu\nu} ,
\label{matching}
\end{equation}
where $t_r$ is the Dynkin index of the multiplet, which is $1/2$ for the fundamental representation,
and $b$ equals  $2/3$ or  $1/6$ respectively for a Dirac fermion and a complex scalar. Indeed one quick way of deriving Eq.~(\ref{matching}) is to require $1/g_{eff}^2$ to satisfy the renormalization group equation of the low-energy theory. In the presence of several multiplets, by diagonalizing the mass matrix, the one-loop effective Lagrangian is simply
\begin{equation}
\frac{g_s^2}{16\pi^2}\left (\frac{2}{3}\sum_{r_F} t_{r_F} \log m_{r_F}(h)+\frac{1}{6}\sum_{r_S} t_{r_S} \log m_{r_S}(h)\right ) G_{\mu\nu}G^{\mu\nu} ,
\label{generalhgg}
\end{equation}
where the two sums are over fermions and scalars, respectively.
 Notice that the contribution of particles with a given spin and color representation is determined by the determinant of the corresponding mass matrix.
 Two different possibilities
can be envisaged for the behavior of the mass eigenvalues on $h$. The first corresponds to multiplets, like the top quark, that are chiral with respect to $SU(2)_L\times U(1)_Y$ in which case $M(h)\propto h$. The corresponding $SU(2)_L$ invariant operator has the form ${\cal O}_{lg}\equiv \log (H^\dagger H) G_{\mu\nu}^a G^{a\,\mu\nu}$, whose singularity at $H=0$ indicates the presence of a massless state. The second case corresponds to vector-like multiplets whose mass does not vanish in the $h\to 0$ limit. The corresponding effective operator should not be singular at $H^\dagger H\to 0$: at lowest order
this corresponds to the ${\cal O}_g$ operator  in Eq.~(\ref{L_SILH}). The on-shell coupling contributing to $h \to gg$ is obtained by making the substitution, $h\to v+h$, in Eq.~(\ref{generalhgg}) and keeping the term linear in $h$
 \begin{equation}
\label{generalcoupling}
{\cal L}_{hgg}=\frac{g_s^2}{48\pi^2}\frac{h}{v}\left (2 \sum_{r_F} t_{r_F} \frac {\partial\log m_{r_F}(v)}{\partial\log v}+\frac{1}{2}\sum_{r_S} t_{r_S}\frac {\partial\log m_{r_S}(v)}{\partial\log v}\right )G_{\mu\nu}^a G^{a\, \mu\nu},
\end{equation}
where we have assumed the scalar Higgs $h$ has a canonical kinetic term. In the presence of ${\cal O}_{H}$, a rescaling $h\to h/\sqrt{1+c_{H} v^2/f^{2}}$ is required
to bring the Higgs kinetic term back to canonical normalization.
Notice that, by $SU(2)_L$ invariance, the total contribution of the heavy multiplets should start at ${\cal O}(v^2)$, corresponding to the ${\cal O}_g$ operator.\footnote{ This result can be directly established by considering the loops with propagating heavy particles and noticing that only the insertion of an even number of external $H$ legs is allowed by $SU(2)_L$ invariance. Equivalently this follows from Eq.~(\ref{generalhgg}) by noticing that  the determinant of the heavy field  mass matrix is an $SU(2)_L$ invariant polynomial of  $H$.}  
The exact one-loop computation parametrized by the function $I_g(m_h^2/m_t^2)$ reduces to the first term inside bracket in the $m_h/m_t\to \infty $ limit. 

We have seen from Eq.~(\ref{generalcoupling}) that the corrections to the Higgs coupling to gluons are determined by the $h$ dependence  of the determinant
of the mass matrix ${\cal M}(h)$ for each particle species
\beq
\sum_{r_F} t_{r_F} \frac{\partial}{\partial \log v} \log \left({\cal M}_{r_F}^\dagger(h) {\cal M}_{r_F}(h)\right) +
\frac14 \sum_{r_S} t_{r_S} \frac{\partial}{\partial \log v} \log \left({\cal M}_{r_S}^\dagger(h) {\cal M}_{r_S}(h)\right).
\eeq
On the other hand,
the one-loop quadratic divergence in the Higgs mass can be computed using 
 the Coleman-Weinberg potential \cite{Coleman:1973jx} and is proportional to
\begin{equation}
\label{cw1}
\frac1{16\pi^2}\, \Lambda^2\, {\rm Str}\, {\cal M}^\dagger(h) {\cal M}(h),
\end{equation}
where we have used the supertrace notation to incorporate both the bosonic and fermionic states. If the
coefficient of the $h^2$ term in Eq.~(\ref{cw1}) is non-vanishing, the Higgs quadratic divergence is not canceled
at the one-loop order. We will
demonstrate that cancellations of the quadratic divergence in Eq.~(\ref{cw1}), or the absence
thereof, allows us to make statements on the sign of $c_g$ induced by integrating
out the new heavy states.\footnote{If the new heavy state is also chiral under $SU(2)_L\times U(1)_Y$, the operator
induced is ${\cal O}_{lg}$. In this case the quadratic divergence
in the top sector is {\em always} enhanced and so is the on-shell coupling of the Higgs to gluons.
The interplay between ${\cal O}_g$
and ${\cal O}_{lg}$ in the di-Higgs production was studied in Ref.~\cite{Pierce:2006dh}.}

\subsection{\label{topsect}The General Fermion Mass Matrix}

We will assume the top quadratic divergence is canceled by 
a new pair of
vector-like colored quark which we call the top partner. 
As discussed in the previous subsection, the operator ${\cal O}_g$ induced by integrating out the top and top partner is
determined by the two-by-two fermion mass matrix. 
We use the notation $(u_3, u_3^c)$ for the fermion having the same quantum number as the SM top quark, and $(U_3, U_3^c)$
for the heavy top partner. Obviously $u_3$ must be embedded in a $SU(2)_L$ doublet $q=(u_3, d_3)$, while $u_3^c$ is a $SU(2)_L$ singlet.
On the other hand, $(U_3, U_3^c)$ can be both $SU(2)_L$ singlets, like in the $SU(5)/SO(5)$ littlest Higgs \cite{Arkani-Hamed:2002qy},
or embedded in two different $SU(2)_L$ doublets, such as in the $SO(9)/SO(5)\times SO(4)$ littlest Higgs \cite{Chang:2003zn}.

We will first consider a general fermion mass matrix in the basis $(u_3, U_3)$, before discussing some prototypes appearing in the
literature. After turning on the Higgs field $h$ as a background field, 
the most general mass matrix is
\beq
\label{eq:genmass}
(u_3^c, U_3^c) {\cal M}\left(\begin{array}{c}
                                                      u_3 \\
                                                      U_3
                                                      \end{array}\right) =
(u_3^c, U_3^c)  \left(\begin{array}{cc}
       {\cal M}_{11}(h) &  {\cal M}_{12}(h) \\
       {\cal M}_{21}(h)  &  {\cal M}_{22}(h)
                                \end{array}\right)   
                               \left(\begin{array}{c}
                                                      u_3 \\
                                                      U_3
                                                      \end{array}\right)                                 ,
\eeq                                                                           
where 
\beq
{\cal M}_{11} = \frac{\lambda_1}{\sqrt{2}}h\left(1-c_1 \frac{h^2}{2f^2}\right) +{\cal O}(h^5) ,\qquad
{\cal M}_{22} = \lambda_2 f +{\cal O}(h^2) ,
\eeq
as dictated by the $SU(2)_L$ symmetry. The form of the off-diagonal entries depends on whether $(U_3, U_3^c)$ belongs to 
$SU(2)_L$ doublets or singlets. In the former case a non-zero ${\cal M}_{21}$ starts at zeroth order in $h$ and ${\cal M}_{12}$
 at linear order in $h$, while the latter case can be described by ${\cal M}^{T}$ in the first case.

The mass eigenvalues are obtained by diagonalizing the mass matrix in Eq.~(\ref{eq:genmass}) through the singular
value decomposition
\beq
\label{eq:diagmass}
{\cal M}(h) \to\left(\begin{array}{cc}
       m_t (h) &  0 \\
       0  &   m_T(h)
                                \end{array}\right)  =  {\cal U}(h)^\dagger {\cal M}(h) {\cal V}(h),
\eeq                                
where $ {\cal U}(h)$ and ${\cal V}(h)$ are complex
rotations in the $(u_3^c, U_3^c)$ and $(u_3, U_3)$ sectors, respectively. The $h$-dependent rotation
generates non-oblique operators, $J_{HL\mu}^a J_{\psi}^{a\mu}$ and $J_{RL\mu}^3 J_{\psi}^{Y\mu}$,
from the gauge-kinetic term of the fermions. However, these dimension-six operators involve only the
third generation SM fermion and current experimental bounds are very weak \cite{Berger:2009hi}.

The mass eigenvalues can be expanded in terms of the background field $h$:
\beq
\label{eq:masseigen}
m_t(h) = \frac{\lambda_t}{\sqrt{2}} h\left(1-c_y^{(t)} \frac{h^2}{2f^2}\right) +{\cal O}(h^5),\qquad
m_T(h)=\lambda_T f \left(1-\frac{\lambda_t^2}{4\lambda_T^2}\frac{h^2}{f^2}\right) +{\cal O}(h^4) .
\eeq
The light mass eigenstate, which is taken to be the SM top quark, must become massless in the limit $h\to 0$,
when the determinant of ${\cal M}$ also vanishes. The coefficient of $h^3$ in $m_t(h)$ is a new contribution to 
the operator ${\cal O}_y$ that is non-universal and specific to the non-linearity in the top Yukawa coupling,
unlike those universal contributions to ${\cal O}_y$ discussed previously in Section \ref{sect:OH} which are associated with the non-linearity in
the Goldstone kinetic term. On the other hand, the $h^2$ term in
the heavy mass eigenvalue $m_T(h)$ represents the heavy top partner's contribution to $c_{g}$, as we shall see. The
cancellation of 
Higgs quadratic divergences requires the absence of ${\cal O}(h^2)$ term in ${\rm Tr}({\cal M}^\dagger {\cal M})$:
\beq
m_t(h)^2+m_T(h)^2 = {\rm constant} + {\cal O}(h^4),
\eeq
which fixes the coefficient of $h^2$
term in the heavy mass eigenvalue $m_T(h)$ in Eq.~(\ref{eq:masseigen}). 

Given the mass eigenvalues, we can compute the contribution to ${\cal O}_g$ when integrating out the SM top quark and
the top partner using Eq.~(\ref{generalcoupling}):
\beq
\label{cancel1}
\left.
\frac{g_s^2}{48\pi^2}\frac{h}{v}\times \frac{1}{2} \frac{\partial}{\partial \log  h}  \log \frac{\det { {\cal M}^\dagger {\cal M}}}{\mu^2}\right|_{h=v} 
 = \frac{g_s^2}{48\pi^2}\frac{h}{v}\left(1 - c_y^{(t)}\frac{v^{2}}{f^{2}}- \frac{m_t^2}{m_T^2}\right),
\eeq
where 
the first two terms on the right hand side are the contributions from the light mass eigenstate, taken to be the SM top, and the last term
from the heavy top partner. The presence of $c_y^{(t)}$ is due to the non-linearity in the top Yukawa sector, while the third term, 
after comparing with the SILH Lagrangian in Eq.~(\ref{L_SILH}), is the contribution to $c_g$ from the heavy top partner:
\beq
\label{eq:cg}
c_g = - \frac16\ ,
\eeq
provided we identify $g_\rho$ with $\lambda_T$.
Notice that
the minus sign is a direct consequence of requiring the Higgs quadratic divergences
be canceled in the top sector, while the number 1/6  is essentially fixed by the 
QCD beta function. 
Since $c_g$ is
always negative, the interference is destructive with the SM top in the gluon fusion production
of the Higgs. 
If to the contrary the Higgs quadratic divergence is enhanced by the top partner, 
then there  would be a positive sign in Eq.~(\ref{eq:cg})
and the interference would be constructive. 

In deriving the on-shell coupling of the Higgs with the gluons in Eq.~(\ref{cancel1}), we have ignored two other effects discussed 
in Section \ref{sect:OH}: the operator ${\cal O}_H$, giving rise to non-canonical normalization of the Higgs kinetic term, and
the universal contributions to ${\cal O}_y$, adding to $c_y^{(t)}$ in Eq.~(\ref{eq:masseigen}). Taking them into account we get
\bea
c_H&=&c_H^{(\sigma)}+c_H^{(s)}+c_H^{(v)} , \quad c_y = c_y^{(\sigma)}+c_y^{(s)}+c_y^{(v)} + c_y^{(t)}, \\
\label{eq:hggexp}
 {\cal L}_{hgg} &=&  \frac{g_s^2}{48\pi^2}\frac{h}{v}\left[1 - \left(\frac{c_H}2+ c_y-\frac{3c_g \lambda_t^2}{\lambda_T^2}\right)\frac{v^{2}}{f^{2}}\right]G_{\mu\nu}^a G^{a\, \mu\nu}  \ .
\eea
Eq.~(\ref{eq:hggexp}) is consistent with Eq.~(\ref{eq:onshellhgg}) once we use the asymptotic value of the form factor $I_g\to -2/3$ for $m_t^2\gg m_h^2$. 

\subsection{Prototypes of Fermion Mass Matrices}
\label{prototypes}
Here we discuss some prototypical fermion mass matrices appearing in the literature of composite Higgs models,
using the same notation as in the original literature.\footnote{We only consider models with one Higgs doublet. In addition, our normalization of generators is such that ${\rm Tr}(X^a X^b)=\delta^{ab}$.} Here we do not include 
the effects of $c_H$ and $c_y$ discussed in Section  \ref{sect:OH}.
\begin{itemize}

\item The $SU(5)/SO(5)$ littlest Higgs model \cite{Arkani-Hamed:2002qy}:
 \beq
 \label{eq:topsu5}
 f (u_3^{c\prime}, \tilde{t}^c ) \left(
\begin{array}{cc}
 -i \sqrt{2} \lambda_1  \sin \left(\sqrt 2 h/f\right)&\lambda_1 \left(\cos \left(\sqrt 2 h/f\right)+1\right) \\
 0 & \lambda_2 
\end{array}
\right)
              \left(\begin{array}{c}
              t \\
              \tilde{t}
              \end{array}\right).
     \eeq

\item The $SO(9)/SO(5)\times SO(4)$ littlest Higgs with custodial symmetry \cite{Chang:2003zn}:
 \beq
 f  (t^{c}, \tilde{t}^c ) \left(
\begin{array}{cc}
 \frac{i}2 y_1 \sin \left(\sqrt 2 h/f\right)& 0 \\
 \frac14 y_1 \sqrt{1+\cos^4 \left( h/\sqrt{2}f\right)} & y_2 
\end{array}
\right)
              \left(\begin{array}{c}
              t \\
              \tilde{t}
              \end{array}\right),
     \eeq
where 
\bea
\tilde{t}&=& \frac1{ \sqrt{1+\cos^4 \left( h/\sqrt{2}f\right)}} \left( -i \cos^2(h/\sqrt{2}f) \chi_{11} + \chi_{22}\right),\nonumber \\
\tilde{t}^c&=& \frac1{ \sqrt{1+\cos^4 \left( h/\sqrt{2}f\right)}} \left( i \cos^2(h/\sqrt{2}f) \chi_{11}^c + \chi_{22}^c\right).
\eea

\item The T-parity invariant $SU(5)/SO(5)$ model with a T-odd top partner \cite{Cheng:2005as}:
 \beq
 \label{eq:tsu5}
 f (U_a^c, U_b^c ) \left(
\begin{array}{cc}
 -i \sqrt{2} \lambda \sin \left(\sqrt 2 h/f\right)&\lambda \left(\cos \left(\sqrt 2 h/f\right)+1\right) \\
  -i \sqrt{2} \lambda \sin \left(\sqrt 2 h/f\right)&-\lambda \left(\cos \left(\sqrt 2 h/f\right)+1\right)
\end{array}
\right)
              \left(\begin{array}{c}
              t \\
              \tilde{t}
              \end{array}\right),
     \eeq
where $t= (t_1+t_2)/\sqrt{2}$ is the T-even SM top quark and $ \tilde{t}=(U_a-U_b)/\sqrt{2}$ is the T-odd top partner.
Furthermore, under T-parity we have $U_a^c\leftrightarrow U_b^c$.   

Given these assignments of T-parity, in 
the mass matrix the second row is simply the image of the first row under T-parity. It is also worth commenting that the above
mass matrix could be diagonalized by going to the T-eigenbasis for the $(U_a^c, U_b^c)$ fermions, which requires
no $h$-dependent rotation. Hence there is no generation of additional non-oblique operators commented below Eq.~(\ref{eq:diagmass}).

\item The toy $SU(3)/SU(2)$ model \cite{Perelstein:2003wd}:
 \beq
 f  (\bar{u}_R, \bar{U}_R ) \left(\begin{array}{cc}
             \lambda_1 i \sin\left(\sqrt 2 h/f\right)  & \lambda_1 \cos\left(\sqrt 2 h/f\right) \\
              0 & \lambda_2 
              \end{array}\right)
              \left(\begin{array}{c}
              u_L \\
              U_L
              \end{array}\right).
     \eeq

\item The T-parity invariant $SU(3)/SU(2)$ model with a T-odd partner  \cite{Cheng:2005as}: 
 \beq
 f  (U_a^c, U_b^c ) \left(\begin{array}{cc}
             \lambda i \sin\left(\sqrt 2 h/f\right)  & \lambda \cos\left(\sqrt 2 h/f\right) \\
             \lambda i \sin\left(\sqrt 2 h/f\right)  & -\lambda \cos\left(\sqrt 2 h/f\right)
              \end{array}\right)
              \left(\begin{array}{c}
              u_3 \\
              T
              \end{array}\right),
     \eeq
     where, similar to Eq.~(\ref{eq:tsu5}), $u_3$ is the T-even SM top quark, $T$ is the T-odd top partner,
     and $U_a^c\leftrightarrow U_b^c$ under T-parity.


\end{itemize}

\subsection{A Remark on $c_y^{(t)}$}
\label{sect:cyobs}

Given the prototypical mass matrices in the previous subsection, one can compute $c_y^{(t)}$ defined in Eq.~(\ref{eq:masseigen})
by computing the mass eigenvalues for the SM top quark. In all cases we find
\beq 
\label{eq:cynoproof}
c_y^{(t)} >0.
\eeq
Loosely speaking, this has to do with the fact that in a nl$\sigma$m the PNGB plays the role of an angular variable in 
$\exp(i h X_h)$, resulting in the oscillatory dependence of the Yukawa coupling on the background Higgs field, as well as the particular implementation of Yukawa couplings. However, we do not find a group-theoretic proof
for Eq.~(\ref{eq:cynoproof}) and will merely take it as an empirical observation. In addition, we observe the same oscillatory dependence of the Yukawa coupling on the Higgs field even in little Higgs models with two Higgs doublets, such as in Refs.~\cite{Chang:2003un, ArkaniHamed:2002qx, Low:2002ws, Skiba:2003yf}, and extra-dimensional models \cite{Barbieri:2000vh}. The same pattern arises in the holographich composite Higgs, to which we devote the next subsection.  

\subsection{Holographic Composite Higgs}

 In holographic composite Higgs models \cite{Agashe:2004rs, Contino:2006qr} one cannot reduce the analysis to a $2 \times 2$ mass matrix in the charge 2/3 quark sector like it was done in section \ref{prototypes}. Not only does the whole KK tower of charge 2/3 fermions contribute to the generation of the Higgs potential (and thus to the finiteness of its mass), but also the resonances in other charge channels are relevant.  In particular in all models, at least, the charge $(-1/3)$ resonances must be included. The expressions for the fermion masses that are given in the literature correspond already to the light eigenvalue, or, equivalently, they describe the low energy Yukawa after having integrated out the KK modes. The results for $c_y$ was already discussed in Ref.~\cite{Giudice:2007fh}. In the model where matter sits in the $\bf 4$ and $\bf 5$ of $SO(5)$ one has respectively $m_f\propto \sin h/f$ and 
$m_f\propto \sin 2h/f$ for  fermion masses, including the top quark.
Again this corresponds to a positive direct contribution $c_y^{(t)}$ in both models. On the other hand when adding $c_y^{(\sigma)}$, which we already saw (see section \ref{subsec:nlsm}) is negative, one finds $c_y=0$ and $c_y=1$, in respectively the model with $\bf 4$ and $\bf 5$ representations \cite{Giudice:2007fh}. Therefore $c_y$ is  never negative.


\subsection{Comments on $c_g$}
\label{commentscg}
We have so far established the sign of $c_g$ induced by the heavy top partner to be negative, if the top quadratic divergence
is canceled, which is the case for all the little Higgs models. However, there are models with top partners in which the top
quadratic divergence is not canceled, such as the universal extra-dimensional models (UEDs) \cite{Appelquist:2000nn}.
In UEDs each SM chiral fermion gets a KK tower of
Dirac fermion which includes both left- and right-handed modes. In this case the partners $Q_n$
of the left-handed top  are electroweak doublets whereas the right-handed partners $T_n$ are singlets. The KK masses are
\begin{equation}
m_t^2=\frac12 (\lambda_t h)^2,\qquad\qquad m_{Q_n}^2=m_{T_n}^2=\frac{n^2}{R^2}   + \frac12(\lambda_t h)^2,\qquad 
n=1,2,3,\cdots \qquad,
\end{equation}
where $R$ is the scale of compactification and could be 600 GeV or lower \cite{Gogoladze:2006br}. 
Comparing to Eq.~(\ref{cancel1}) it is thus evident that the KK modes enhance the Higgs quadratic divergence in the top sector and interfere positively
with the top quark contribution. 

Since ${\cal O}_y$ and ${\cal O}_H$ also contribute to the on-shell amplitude of the Higgs with the gluons,
it is worth emphasizing that
only under circumstances where effects of  ${\cal O}_y$ and ${\cal O}_H$ are small can
we directly infer the magnitude of the cross-section from the sign in front of ${\cal O}_g$. 
In the UEDs model considered in Ref.~\cite{Petriello:2002uu}, the five-dimensional Yukawa coupling is identical
to the four-dimensional Yukawa coupling and hence $c_y^{(t)}=0$. On the other hand, because of KK parity the coupling
of the first KK gauge boson (as well as all the odd KK-mode) with the SM Higgs is forbidden at the tree level, while
the effect from the second KK mode is suppressed by the approximate KK-number conservation. Thus there is 
no ${\cal O}_H$ and ${\cal O}_y$ induced from integrating out the KK gauge bosons at leading order. Furthermore, no
triplet or singlet scalars are present in the model. In the end we expect $\Gamma(gg\to h)$ to be dominated by the effect
from ${\cal O}_g$ and be enhanced relative to the SM rate, which is consistent with the explicit computation in 
Ref.~\cite{Petriello:2002uu}. Our analysis indicates 
that this increase is a direct consequence
of the absence of cancellation mechanisms in the Higgs quadratic divergences. The same arguments would apply to the 
case of warped KK-parity \cite{Agashe:2007jb}  as well.

While the sign correlation in the corrections to $m_h^2$ and $c_g$ that we just pointed out works in the simplest models of interest, it does not work in the most general situation with more than one heavy partner of the top. Consider indeed generalizing the $2\times 2$ mass
matrix to the $3\times 3$ case
\begin{equation}
\label{mass3}
{\cal M}^\dagger {\cal M} = \left(
     \begin{array}{ccc}
       (\lambda_t h)^2/2 & 0&0 \\
       0 & \lambda_{2T}^2 f^2 \left[1 +  c_2 (h/f)^2 \right]^2&0\\
       0&0& \lambda_{3T}^2 f^2 \left[1 +  c_3 (h/f)^2 \right]^2
      \end{array}
      \right).
\end{equation}
The cancellation of the quadratic divergence reads
\begin{equation}
\label{natural3}
4\left (\lambda_{2T}^2c_2+\lambda_{3T}^2c_3\right )+\lambda_t^2=0 ,
\end{equation}
while 
\begin{equation}
{\rm Tr}\log {\cal M}^\dagger {\cal M}= \log h^2+\left  (c_2+c_3\right )\frac{h^2}{f^2}+ {\cal O}(h^4) \ .
\end{equation}
It is clear that, without violating Eq.~(\ref{natural3}), one can choose $c_2+c_3>0$ which corresponds to an overall 
constructive interference with the SM top contribution to $h\to gg$. Nonetheless the parameter choice to realize that seems a bit contrived and we are also not aware of interesting models that realize it. Indeed in the holographic composite models, where a whole tower of KK models controls the Higgs potential, this contrived choice does not arise and the coefficient $c_g$ turns out negative \cite{Falkowski:2007hz}. It should however be noticed, as we discuss in the next section, that in the preferred  region of parameters of those models the effects of $c_g$ are subleading to those of $c_H$ and $c_y$ \cite{Giudice:2007fh}.

\subsection{Naturalness Constraints on ${\cal O}_\gamma$}
\label{sect:cgamma}

It is clear from the discussion in Section \ref{sub:higgs} that the SM top also makes a contribution to the effective Higgs-photon coupling
$(h/v)(F_{\mu\nu})^2$ given by its contribution to the one-loop beta function of QED. However, recall that
all electrically charged particles with a 
significant coupling to the Higgs will also induce the effective Higgs-photon coupling, in particular the SM gauge boson $W^\pm$
\cite{Kniehl:1995tn}:
\beq
\label{eq:hgaga}
{\cal L}_{eff} = -\frac14\left( \frac1{e^2(\mu)} - 7\times \frac{1}{8\pi^2} \log\frac{m_W(h)}{\mu}
       + \frac43\times \frac{1}{8\pi^2} \log\frac{m_t(h)}{\mu}\right) F_{\mu\nu} F^{\mu\nu},
\eeq
where we have included both the $W$ boson and the top quark contributions, whose coefficients are $-7$ and $+4/3$ respectively.
Observe that the $W$ boson contribution is dominant and several times larger than the top contribution. 
In addition, the signs are opposite
in the two contributions. 

If we consider only contributions from a new set of heavy electroweak gauge bosons and 
neglect the fermions, then it is possible to make a connection between the sign of ${\cal O}_\gamma$ and the cancellation of 
Higgs quadratic divergence in the electroweak sector, using arguments completely parallel to that in the case of ${\cal O}_g$.
In the end,
similar to the case of Higgs-gluon coupling, naturalness consideration requires the heavy and light gauge bosons to interfere
destructively in the partial decay width $h\to \gamma\gamma$, if there is cancellation of Higgs quadratic divergences. However, the effect is
suppressed by $m_W^2/m_{W'}^2$ and smaller than the SM one-loop effect, given the electroweak constraint on a heavy $W'$, unless there
 is a symmetry such as the T-parity.


\subsection{Supersymmetry}
\label{section:susy}

At last we briefly consider the constraint on ${\cal O}_g$ in supersymmetry, where the top partners are spin-zero particles.
It has been discussed in the literature \cite{Djouadi:1998az, Carena:2002qg, Dermisek:2007fi} that the top squark contribution 
to the gluon fusion production of the lightest CP-even Higgs interferes constructively with the SM top contribution, 
when the off-diagonal mixing
term is small in the top squark mass matrix, and destructively when the mixing is large. Here we will discuss the connection
of this pattern with the cancellation of top quadratic divergence by the top squark.

Using the notation $(\tilde{t}_L^\dagger, \tilde{t}_L)$
and $(\tilde{t}_R^\dagger,\tilde{t}_R)$ in the electroweak eigenbasis.,
the stop contribution to the one-loop beta function of QCD is
\beq
\label{eq:stophgg}
{\cal L}_{eff} = -\frac14\left( \frac1{g_s^2(\mu)} - \frac23 \times \frac{1}{8\pi^2} \log\frac{m_t(h)}{\mu}
       - \frac1{12}\times \frac{1}{8\pi^2} \log\frac{{\cal M}^\dagger_{\tilde{t}}{\cal M}_{\tilde{t}}}{\mu^2}\right) G^a_{\mu\nu} G^{a\,\mu\nu}, 
\eeq
where ${\cal M}^\dagger_{\tilde{t}} {\cal M}_{\tilde{t}}$ is the mass matrix-squared of the top squarks. 
Then the interference effect of the scalars can be determined by looking at the $h$ dependence
of $\det {\cal M}_{\tilde{t}}^\dagger {\cal M}_{\tilde{t}}$, similar to the case with a fermionic top partner.
On the other hand, the cancellation of quadratic divergences is dictated by the supertrace
 ${\rm Str}\ {\cal M}^\dagger {\cal M}$.

When there is no
mixing in the top squark sector, both $\tilde{t}_L$ and $\tilde{t}_R$ contribute equally to the cancellation of 
top quadratic divergence, resulting in the mass matrix:
\begin{equation}
{\cal M}^\dagger_{\tilde{t}} {\cal M}_{\tilde{t}} = \left( \begin{array}{cc}
      \tilde{m}_L^2 + (\lambda_t h)^2/2  & 0 \\
     0 & \tilde{m}_R^2 + (\lambda_t h)^2/2 
          \end{array} \right),
\end{equation}
where $\tilde{m}_{L,R}^2$ are the soft-breaking masses. Notice that,
 because of the spin-statistics, there is a crucial difference in signs when comparing
with Eq.~(\ref{eq:masseigen}). The induced ${\cal O}_g$ is given by
\begin{eqnarray}
\label{cancel3}
&& \left.\frac{\partial}{\partial \log h} 
   \log \frac{m_t(h)}{\mu}\right|_{h=v} 
 +  \frac1{8}\left.\frac{\partial}{\partial \log h} 
   \log \frac{\det {\cal M}^\dagger_{\tilde{t}} {\cal M}_{\tilde{t}}}{\mu^2} \right|_{h=v} 
       \nonumber \\
&& = 1 + \frac14\left( \frac{m_t^2}{2\tilde{m}_L^2 + m_t^2/2}
         +  \frac{m_t^2}{2\tilde{m}_R^2 + m_t^2/2}\right) ,
\end{eqnarray} 
where we see that, opposite to the case of a fermionic top partner, the interference is
constructive when there is cancellation in the Higgs quadratic divergence. 
This difference in signs traces its origin to the fact that fermions and scalars contribute
with opposite sign in the cancellation of Higgs divergences, ${\rm Str} {\cal M}^\dagger{\cal M}$.

Notice that when the mixing in the top squark sector is introduced, the top squark mass determinant
in Eq.~(\ref{cancel3})
is reduced by the off-diagonal mixing term. If the mixing term is large enough, the interference between
the SM top quark and the top squarks could become destructive. Such a behavior
in the minimally supersymmetric standard model has been observed in Refs.~\cite{Djouadi:1998az, Carena:2002qg, Dermisek:2007fi}.
Another example where the off-diagonal mixing term in the top squark mass matrix turns the interference effect from
being constructive to destructive is the five-dimensional supersymmetric model of Ref.~\cite{Barbieri:2000vh}, as was
studied in Ref.~\cite{Cacciapaglia:2001nz}. In general the fact that there is no strict sign correlation for $c_g$ in supersymmetry simply follows from the presence of at least two new mass eigenvalues in addition to the top quark (see discussion at the end of subsection \ref{commentscg}).

\section{Discussions and Summary}
\label{sect:discussion}

In the section we will present a synthesis of the results scattered through the paper and summarize.

\subsection{Size of Effects}

One first issue we must clarify concerns the expected size and nature of the effects. This is useful  in order to compare them to similar effects  arising at higher-loop order in the SM. We first consider composite Higgs models for which the SILH power counting in Section \ref{sect:powercount} applies.
There are two classes of effects:
the first class,  corresponding to $c_H$ and $c_y$,  are ${\cal O}(v^2/f^2)$ and sensitive to the underlying strong dynamics giving rise to the composite nature of the Higgs, while effects in the second class,  $c_g$ and $c_\gamma$,  are
parametrically ${\cal O}(g_{SM}^2v^2/g_\rho^2 f^2)$.\footnote{Recall that the SM contributions to $c_g$ and $c_\gamma$ start at one-loop order.}
Notice that by the last estimate we indicate collectively both $m_t^2/m_T^2$ in the top sector and $m_W^2/m_{W'}^2$ in the vector sector, which in given models could be quantitatively quite different. In fact we know that in little Higgs model $(m_t/m_T)^2$ is preferred to be large, by naturalness considerations, while 
$m_W^2/m_{W'}^2$ is preferred to be small by the constraint on the $S$ 
parameter.\footnote{See the discussion in Section 3
of Ref.~\cite{Giudice:2007fh}.}

We will consider the expected size of $v^2/f^2$ in two broadly defined scenarios: the composite Higgs model without the 
collective breaking mechanism (nCB), which includes the original Georgi-Kaplan model and 
the holographic Higgs model, and the little
Higgs models (LH) which implement the collective breaking. 
The two scenarios differ in the way they engineer a hierarchy in $v/f$:
\beq
{\rm nCB}: \frac{v^2}{f^2}={\cal O}(1)\times \epsilon_{nCB},\qquad{\rm LH}: \frac{v^2}{f^2}={\cal O}\left(\frac{g_\rho^2}{16\pi^2}\right) \times \epsilon_{LH},
\eeq
where the $\epsilon$'s quantify  the amount of tuning in electroweak symmetry breaking. In the absence of fine-tuning, $\epsilon
={\cal O}(1)$, there is no separation in the two scales $v$ and $f$ in the nCB models while a separation of one-loop order is achieved
in LH theories via the collective breaking mechanism. In this case the $v^2/f^2$ effect in LH is always smaller than in
the nCB models by a loop factor, unless $g_\rho\sim 4\pi$. (However in this case the Higgs mass in LH is fine-tuned and collective
breaking does not work.) The situation changes when one considers precision electroweak constraints such as the $S$ parameter \cite{Giudice:2007fh}:
\beq
\hat{S} \alt 2\times 10^{-3} \sim \frac{g_{SM}^2}{16\pi^2} .
\eeq
The contribution to $\hat{S}$ in both nCB and LH models, barring T-parity, is given by
\beq
\hat{S} \sim \frac{m_W^2}{m_\rho^2} = \frac{g_{SM}^2}{g_\rho^2} \times \frac{v^2}{f^2}.
\eeq
Therefore the $S$ parameter requires
\beq
\label{eq:voverf}
\frac{v^2}{f^2} \alt \frac{g_\rho^2}{g_{SM}^2} \times \frac{g_{SM}^2}{16\pi^2} = \frac{g_\rho^2}{16\pi^2},
\eeq
which can 
 be achieved in LH theories without fine-tuning, $\epsilon_{LH}\sim {\cal O}(1)$, while in nCB the amount of fine-tuning
necessary is $g_\rho^2/16\pi^2$.

The relative importance of the ${\cal O}(v^2/f^2)$ effects from ${\cal O}_H$ and ${\cal O}_y$ 
 is determined by comparing Eq.~(\ref{eq:voverf}) with the SM one-loop effect, which is ${\cal O}(g_{SM}^2/16\pi^2)$.
 In nCB models we need $g_{\rho}\gg g_{SM}$, otherwise the Higgs quartic coupling $\lambda\sim g_{SM}^2 g_\rho^2/16\pi^2$ is 
too small and the Higgs mass is below the direct experimental lower bound.\footnote{Such a small quartic was the fatal problem of the first attempts to gauge-Higgs unification in extra-dimensional theories, see for instance a discussion in Ref.~\cite{Scrucca:2003ra}.} In LH, instead, the quartic coupling arises at tree level $\lambda\sim g_{SM}^2$ and thus the lower bound on the Higgs mass does not constrain $g_\rho$, which can be as small as $\sim g_{SM}$. 
 Of course in both nCB and LH the constraint $g_\rho \alt 4 \pi$ is required to have
 some calculability. But notice also that in nCB models bigger $g_\rho$ means less tuning is needed to satisfy Eq.~(\ref{eq:voverf}). In the end, $v^2/f^2$
  in nCB models  is expected to be much larger that the typical SM one-loop effect, while in LH models $v^2/f^2$ can be as small as the typical SM one-loop effect. 
However, as discussed in Ref.~\cite{Giudice:2007fh}, precision electroweak constraints do not prefer
 such a low value of $v^2/f^2$ in LH, which can conceivably be slightly bigger without making the LH model less plausible (that is more fine-tuned).  The only LH model where $g_\rho \sim g_{SM}$ is mandated is the one with T-parity, in which case 
precision electroweak constraints are avoided due to T-parity. 
We conclude that, in a significant region of parameter space of nCB and LH, the effects associated to $c_H$ and $c_y$ 
 do stand out over the `background' of ordinary one-loop effects within the SM.
 Especially for nCB models where $v^2/f^2$ is favored large, effects of $c_H$ and $c_y$ have a chance of being detectable at the  LHC. Needless to say, they are  surely of interest for precision measurements at a linear collider working as the Higgs factory.

Now we discuss the second class of effects, focusing on ${\cal O}_g$. As mentioned above, the size of the effect here comparing
to the SM contribution is
\beq
\frac{m_t^2}{m_T^2} = \frac{\lambda_t^2}{\lambda_T^2} \times \frac{v^2}{f^2}.
\eeq
So, even without fine-tuning, in LH models this effect is formally one SM-loop smaller than the SM top effect, i.e. it is of the order of NLO effects in the SM. In nCB theories,
given the fine-tuning necessary to satisfy the $S$ parameter, this is also suppressed by a SM loop factor, without the $(g_\rho/g_{SM})^2$ enhancement seen in the first class of effect. However, if the Higgs is naturally light and close to the LEP lower bound, the top quadratic
divergence in the Higgs mass is cut off by the top partner,
\beq
\delta m_H^2 \sim \frac{3\lambda_t^2}{8\pi^2} m_T^2,
\eeq
which implies $m_T$ is only a few hundreds GeV. Therefore, it is important to stress that, although formally
a loop factor,
$m_t^2/m_T^2$ could end up being numerically important.

It is finally worth mentioning  a class of models for which the SILH power counting does not apply because of symmetry reasons and the (more) fundamental nature of the Higgs. This class includes models with a fundamental Higgs scalar and a new parity at the TeV scale, such as the R-parity in supersymmetry and KK-parity in extra-dimension \cite{Appelquist:2000nn}.\footnote{KK parity models have normally a cut-off exceeding 10 TeV, and thus the Higgs can be taken as elementary to a good approximation.} In this case the Higgs sector does not belong to a nl$\sigma$m (below at least tens of TeV) and 
the only contributions to $c_H$ and $c_y$ come from integrating out heavy vectors and scalars, which can only be loop-induced due to the new parity. Then $c_H$ and $c_y$ are in the order of $(g_{SM}^2/16\pi^2)\times (v^2/f^2)$, where we have taken 
the typical coupling strength associated with the heavy particle to be $g_{SM}$, as is the case in existing literature.
Furthermore,
 the scale $f$ is defined as the ratio of the heavy mass $m_{new}$ with $g_{SM}$. As such, these effects are much smaller than the SM one-loop effect. On the other
hand, the effects in $c_g$ and $c_\gamma$ are still suppressed by $m_{SM}^2/m_{new}^2$, which could turn out to be numerically important if $m_{new}$ is not too far from the mass of the SM particles associated in the process.

We summarize the discussion above as follows:

\begin{itemize}

\item The effects of $c_H$ and $c_y$ can stand out over SM loop effects for nCB and LH models with $g_\rho \gg g_{SM}$. On the other hand, for the same models, the effects of $c_g$ and $c_\gamma$ are formally of order  $\sim g_{SM}^2/16\pi^2$ with respect to the leading SM contribution.  

\item In the main stream choice of LH parameters where $g_\rho \sim g_{SM}$ and for the T-parity models\footnote{With T-parity, even though there is no tree level effect from integrating out heavy states, there are still contributions to $c_H$ and $c_y$ from the nl$\sigma$m in the UV.} the effects of $c_H$, $c_y$, $c_g$, and $c_\gamma$ are all formally of the order of NLO contributions in the SM.


\item In models that do not address the little hierarchy problem (such as UED) the effects of $c_H$ and $c_y$ can  even be smaller than $g_{SM}^2/16\pi^2$, while $c_g$ and $c_\gamma$ can be enhanced over $g_{SM}^2/16\pi^2$
due to the lightness of new heavy states. In this scenario the Higgs is a fundamental scalar up to at least tens of TeV and there is a new parity at the TeV scale to suppress tree level effects from integrating out heavy particles.

\end{itemize}

A final word of caution: when we classify effects as being  $\sim$ one-loop in SM, it is important to keep in mind that
in the SM there is indeed a wide spectrum of one-loop effects ranging from $\alpha_W/4\pi \sim 0.1\%$, appearing in electroweak precision quantities, to $3\lambda_t^2/8\pi^2\sim 5 \%$, controlling threshold corrections to the Higgs mass from heavy states.
 Our estimate of the size of the effects has to be taken {\it cum grano salis} as a  rough guideline.

Let us now synthesize the information we obtained on the signs of $c_H$ and $c_y$, which are scattered in previous sections.

\subsection{Sign of $c_H$}

In the limit where the SM gauge couplings are turned off, $c_H$ can be related 
to UV physics via a dispersion relation in Eq.~(\ref{Dispersion relation}). The presence of two contributions of opposite sign
in the dispersive integral suggests that, unlike other forward amplitudes, no
general positivity theorem can be proven  for $c_H$. In addition to deriving the dispersion relation, we explicitly computed
three leading sources for $c_H$: the genuine nl$\sigma$m contribution $c_H^{(\sigma)}$ in Eq.~(\ref{eq:chnlsm}),  and the contributions from exchanges of  scalars and vectors, $c_H^{(s)}$ in Eq.~(\ref{totalscalar}) and $c_H^{(v)}$ in Eq.~(\ref{eq:chinvec}), respectively.
We found that $c_H^{(\sigma)}$ and $c_H^{(v)}$ are always positive while $c_H^{(s)}$ is negative when there are triplet scalars, which nicely fits with the expectation from the dispersion relation that $c_H$ could be negative only in the presence of a doubly-charged state.

Despite a negative $c_H$ from triplet scalars, by a detailed study of the phenomenologically relevant cases, we discovered that in the
end the total $c_H$ resiliently comes out
positive. In fact we present below a proof that $c_H$ is positive in a general  little Higgs type of theories.

First it is useful to recall the general structure of composite Higgs models.  
At the level of  the microscopic theory above the scale $f$ and neglecting $g_\rho$,
one has a nl$\sigma$m based on the coset  $G/H$ to which the Higgs belongs. Neglecting the existence of fermions for now,
when $g_\rho$ is turned on with $g_{SM}=0$ the coset is reduced to $G_{IR}/H_{IR}$, which means that by integrating out the scalars and the vectors at the scale $m_\rho$ the Higgs sector is still {\it exactly} described by a $G_{IR}/H_{IR}$ nl$\sigma$m.\footnote{More specifically, $G_{IR}$ ($H_{IR}$) can be generators in $G$ ($H$) which commute with the  generators gauged with strength $g_\rho$.} Introducing the fermions would not spoil the $G_{IR}/H_{IR}$ coset structure as long as the fermion is
only charged under $G_{IR}$. In this case,
we could immediately conclude that  the value of $c_H$ below $m_\rho$ must flow
to that of a $G_{IR}/H_{IR}$ nl$\sigma$m and be positive, despite the negative contribution from triplet scalars.

After turning on $g_{SM}$ 
the residual $G_{IR}$ is explicitly broken down to the SM gauge group, and 
the Higgs is no longer described by an exact nl$\sigma$m. 
In other words, at the scale below $m_\rho$ the effective theory using 
a $G_{IR}/H_{IR}$ nl$\sigma$m is only valid up to corrections of 
${\cal O}(g_{SM}^2/g_{\rho}^2)$. Thus in the limit $g_\rho \gg g_{SM}$ the infrared coset structure is a good 
approximation of the Higgs sector, while away from this limit one cannot claim any coset structure in the effective Lagrangian below 
$m_\rho$.\footnote{For instance in the $SU(5)/SO(5)$ littlest Higgs where two copies of $SU(2)$ subgroups are gauged, the UV coset is 
$SU(5)/SO(5)$. In the limit in which one gauge group, say $SU(2)_R$, is more strongly coupled 
than the other, $g_R=g_\rho \gg g_L = g_{SM}$, the Higgs sector below the heavy vector and heavy 
scalar mass scale $(m_\rho)$ is described by a $SU(3)/SU(2)_L$ nl$\sigma$m \cite{Giudice:2007fh}, up to effects of order 
$g_{SM}/g_{\rho}$, if the SM fermion is charged only under weakly gauged $SU(2)_L$.}

The above arguments indicate the total contribution to $c_H$ below $m_\rho$ is positive in the limit where ${\cal O}(g_{SM}^2/g_\rho^2)$ corrections
are negligible:
\beq
\lim_{g_\rho \gg g_{SM}} \ c_H = c_H^{(\sigma)}+c_H^{(s)}+c_H^{(v)} \ >\ 0.
\eeq
The only possible negative contribution is in $c_H^{(s)}$, when triplet scalars are present, which
cannot overcome other positive contributions. Now
consider a LH theory without T-parity where the SM fermion is charged only
under the weakly gauged $SU(2)_L$ group. Due to the collective breaking structure, the form of the scalar potential in a general 
LH model is identical to the one in Eq.~(\ref{eq:scalarp}),
where we showed the triplet scalar contribution 
\beq
c_H^{(s)}= - \frac{(g_R^2-g_L^2)^2}{(g_R^2+g_L^2)^2}
\eeq
is maximally negative when $g_\rho \gg g_{SM} $ (or equivalently $g_R\gg g_L$).
On the other hand, we see from Eqs.~(\ref{eq:chnlsm}) and (\ref{eq:chvec}) that
 neither $c_H^{(\sigma)}$ nor $c_H^{(v)}$ is dependent on $g_\rho$ and $g_{SM}$. Then the fact that
the scalar contribution $c_H^{(s)}$ becomes less negative when $g_\rho \sim g_{SM}$ implies
\beq
\lim_{g_\rho \sim g_{SM}} c_H\  >\  \lim_{g_\rho \gg g_{SM}}  c_H \ >\ 0.
\eeq
 We conclude that in all LH models without T-parity $c_H>0$.

The other case of interest is LH theories with T-parity. In this case 
$g_L=g_R$ and the $G_{IR}/H_{IR}$ coset structure is completely lost in the infrared.
However, since
there is no 
tree level interaction between heavy vectors and scalars with the SM matter, 
$c_H^{(s)}=c_H^{(v)}=0$, we conclude $c_H>0$ in LH models with T-parity.

\subsection{Sign of $c_H+2c_y$}

Although $c_y$ modifies the Yukawa coupling of the fermion, in on-shell coupling of the Higgs with the fermion the
physical combination is $c_H+2c_y$. Thus,
instead of making a statement on $c_y$,
here we will establish the positivity of $c_H+2c_y$.
In this case we can identify four classes of effects: a genuine UV contribution in Eq.~(\ref{eq:chnlsm}), a contribution from integrating out
heavy scalars and vectors,  as in Eqs.~(\ref{eq:scalarchcy}) and (\ref{eq:vecchcy}) respectively, and a non-universal contribution
associated with the non-linearity in the Yukawa interaction in Eq.~(\ref{eq:masseigen}). The total contribution is 
\beq
c_H+2c_y = [c_H^{(\sigma)}+2c_y^{(\sigma)}]+[c_H^{(s)}+2c_y^{(s)}]+[c_H^{(v)}+2c_y^{(v)}]+2c_y^{(t)}.
\eeq
We have proven that all three contributions in the square bracket are positive, even though 
$c_y^{(\sigma)}$, $c_H^{(s)}$ (in the presence of
triplet scalars), and $c_y^{(v)}$ are negative.
 Moreover, in Section \ref{sect:cyobs} we remarked that $c_y^{(t)}>0$ in all phenomenologically relevant models we surveyed.
Thus, we conclude in all composite Higgs models that are presently on the market, $c_H+2c_y>0$, implying the Higgs coupling to the quark is reduced.

\subsection{Summary}
\label{sect:conclusion}

In this work we initiated a study on theoretical constraints on Higgs effective couplings. In particular, we consider four dimension-six operators, 
${\cal O}_H$, ${\cal O}_y$, ${\cal O}_g$, and ${\cal O}_\gamma$, which could directly impact the production and decay of the Higgs boson, among others.
The operator ${\cal O}_H$, in addition, affects the scattering of the longitudinal components of the SM $W$ boson.  We have identified several sources for these operators, depending on the nature of the Higgs boson. First, if we assume the Higgs is a fundamental scalar, we find

\begin{itemize}

\item the contribution from integrating out heavy states at tree level satisfies $c_H>0$ except when there are the triplet scalars coupling to the Higgs, implying the
 on-shell
coupling of the Higgs  boson is reduced from the SM expectations if effects from other operators can be neglected.

\item  $c_H+2c_y>0$ when integrating out heavy scalars (including the triplet scalars) and vectors, implying the overall Higgs coupling to fermions are reduced from
the SM expectations.

\item $c_g<0$ if there is a new colored fermion canceling the top quadratic divergence, implying the interference with the SM top is destructive, while
 $c_g>0$ if the new colored fermion add to the top quadratic divergence. Similarly for $c_\gamma$.

\end{itemize} 
Next assuming the Higgs sector is composite and belongs to a nl$\sigma$m, we find

\begin{itemize}

\item $c_H>0$ and $c_H+2c_y > 0$ for the contribution from the non-linearity in the Higgs kinetic term.

\item $c_y>0$ for the contribution from the non-linearity in the Yukawa interaction, among the models we surveyed.

\end{itemize}
Finally assuming the underlying model has a collective breaking mechanism such as in little Higgs theories,

\begin{itemize}

\item $c_H > 0$ even in the presence of triplet scalars.

\end{itemize}

One representative observable for which our analysis could be useful is the production cross section of the Higgs boson in the gluon fusion channel, which is the dominant production mechanism of the Higgs in a hadron collider. In this particular case we are able to make the statement that, in all composite Higgs models, the production cross section is reduced from the SM expectation. As another example, the Higgs coupling to the SM top quarks, relevant for the associated Higgs production with the top quarks, is also reduced from the SM expectation.

The size of the effects from these operators depends on the particular underlying model. In most cases, they are expected to range from modest, in the order 
of $10 - 30$\%, to as small as the SM one-loop effect, in which case it will be impossible to observe them at the LHC \cite{Duhrssen:2004uu}. However, in the event that an enhancement over the SM expectation is observed in the experimental observables mentioned above, our results show 
that all the composite Higgs models will be strongly disfavored. In this case, very likely the naturalness principle fails to work (at least in the top sector), or there exists $SU(2)_L$ triplet scalars giving rise to a large contribution in ${\cal O}_H$.\footnote{To avoid constraints from the $\rho$ parameter the triplet scalars would have to be in the $(\mathbf{3}_L,\mathbf{3}_R)$ representation of $SO(4)\simeq SU(2)_L\times SU(2)_R$.} More radically, there is also the possibility that the scalar being observed is the dilaton instead \cite{riccardo}.

The theoretical results in this study motivate to further improve on both the theoretical and experimental uncertainties 
\cite{Anastasiou:2005pd} in the
extraction of Higgs production rate in the gluon fusion channel at the LHC. On the other hand, in the unlikely situation of a null discovery at the LHC, our work 
suggests that  a wealth of information could still be revealed by precise measurements of the Higgs coupling, which could be achieved in 
 a high energy $e^+e^-$ collider such as the International Linear
Collider or even a photon collider.

\begin{acknowledgments}
This work is supported in part by the U.S.~Department
of Energy under grant DE-AC02-06CH11357 and by the Swiss National Science Foundation under contract
No. 200021-116372. We acknowledge the hospitality of theory groups at CERN, EPFL Lausanne, and Northwestern University during 
the completion of this work. I.L.~ also acknowledges the hospitality of the Aspen Center for Physics where part of this work was completed.

\end{acknowledgments}

\section*{APPENDIX A: $SO(4)$ Generators and Notations}
\label{appendixA}

In this appendix we list the basis of $SO(4)$ generators adopted in this paper. The $SO(4)$ commutation relations are
\begin{equation}
[T^{mn}, T^{op}] = \frac{i}{\sqrt{2}}(\delta^{mo} T^{np}-\delta^{mp}T^{no}+\delta^{np} T^{mo} -\delta^{no}T^{mp}) ,
\end{equation}
where $\{m,n,o,p\}=\{1,2,3,4\}$. There are several ways to decompose the generators into $SU(2)_L\times SU(2)_R$. Our choice
is
\begin{eqnarray}
T_L^A&=&\left\{\frac1{\sqrt{2}}(T^{23}-T^{14}), \frac1{\sqrt{2}}(T^{13}-T^{42}), \frac1{\sqrt{2}}(T^{21}-T^{43})\right\}, \\
T_R^A&=&\left\{\frac1{\sqrt{2}}(T^{32}-T^{14}), \frac1{\sqrt{2}}(T^{31}-T^{42}), \frac1{\sqrt{2}}(T^{21}-T^{34})\right\},
\end{eqnarray}
where $A=\{1,2,3\}$. Then it is straightforward to check that
\begin{equation}
[T_\chi^A, T_{\chi'}^B]=i\epsilon^{ABC} \delta_{\chi\chi'} T_\chi^C ,
\end{equation}
with $\{\chi,\chi'\}=\{L,R\}$. The hypercharge generator $Y=T_R^3$.
Generators in the vector representation of $SO(4)$ is given by
\begin{equation}
(T^{mn})_{op}=-\frac{i}{\sqrt{2}} (\delta^{mo}\delta^{np}-\delta^{no}\delta^{mp}),
\end{equation}
which results in the normalization ${\rm Tr}(T^A T^B)=\delta^{AB}$.

Given the above basis, for an $SO(4)$ vector $\vec{h}=(h^1, h^2, h^3, h^4)^T$, the $SU(2)_L\times U(1)_Y$ representation is
\begin{equation}
\label{eq:defineh}
H=\frac1{\sqrt{2}}\left( \begin{array}{c}
          h^+ \\
          h 
          \end{array} \right) =
          \frac1{\sqrt{2}}\left( \begin{array}{c}
          h^1+i h^2\\
          h^3+i h^4 
          \end{array} \right).
 \end{equation}         
When the neutral component gets a VEV, $\langle h\rangle = v =245$ GeV,  the $SO(4)=SU(2)_L\times SU(2)_R$ is broken down
to the diagonal subgroup which is the custodial $SU(2)_C$.

\section*{APPENDIX B: Integrating Out Heavy Vectors Within the NL$\Sigma$M}
\label{sect:vect_nlsm}

In this appendix we compute effects of integrating out heavy vector fields in a general nl$\sigma$m, and verify explicitly that
$c_H$ contains no effects of the order $g_{SM}/g_\rho$.
As mentioned in Section \ref{sect:powercount}, we would like to use an operator basis
in which all new physics effects, including those from integrating out heavy fields, are oblique. 
Some care must be taken to preserve this operator basis when integrating out heavy vector fields. 
The most straightforward way is to use  
 the gauge field coupling to the SM fermionic current as the interpolating field for the low-energy field; the heavy field
can be any combination that is linearly independent of the low-energy field, for example the heavy mass eigenstate.
This is the interaction basis taken by Refs.~\cite{Barbieri:2004qk,Marandella:2005wd} and will be adopted here. 
Alternatively, one could use the
mass eigenstates for both the heavy and light gauge fields, and then perform a field-redefinition  in the massless gauge field to
eliminate non-oblique operators. It can be verified that such an approach yields identical results.

The scenario of most interest, which we consider here, is gauging two
subgroups ${\cal G}_1\times {\cal G}_2$ of $G$ with corresponding gauge fields $g_1 A_1^\alpha Q_1^\alpha$ and 
$g_2 A_2^\alpha Q_2^\alpha$, which
is broken down to the diagonal subgroup ${\cal G}_v$ with generators $Q^\alpha_v=Q_1^\alpha+Q_2^\alpha$. The broken axial subgroup
${\cal G}_A$ has generators $Q^a_A=Q_1^\alpha-Q_2^\alpha$.
It is convenient to define
\bea
A_G &\equiv& g_1 A_1^\alpha Q_1^\alpha + g_2 A_2^\alpha Q_2^\alpha = A_v^{\alpha} Q_v^{\alpha} + A_A^{\alpha} Q_A^{\alpha},\\
A_v^{\alpha}&=& \frac12(g_1 A_1^{\alpha} + g_2 A_2^{\alpha}), \quad A_A^{\alpha} = \frac12(g_1 A_1^{\alpha} - g_2 A_2^{\alpha})  .
\eea
The Cartan-Maurer one-form is
\beq
U^\dagger (\partial + i A_G ) U =
 i A_v + i A_A + \frac{i}f D_G \Pi + \frac1{2f^2}[\Pi, D_G \Pi] -\frac{i}{6f^3} [\Pi, [\Pi, D_G\Pi]]+{\cal O}(1/f^4) ,
\eeq
where $D_G\Pi\equiv \partial \Pi + i\, [A_G, \Pi]$ is the gauge-covariant derivative of $\Pi$ under the full symmetry
group $G$. In the above we have suppressed 
the Lorentz index to avoid cluttering 
of indices. The Goldstone-covariant derivative is
\beq
i{\cal D}^a X^a = i A_A^\alpha Q_A^\alpha + \frac{i}f D_v \Pi -\frac1f [A_A,\Pi]_X + \frac1{2f^2}[\Pi, D_v\Pi]_X +
 \frac1{2f^2} [\Pi, i[A_A,\Pi]]_X +\cdots \quad .
\eeq
Note that $D_v\Pi=\partial\Pi+i[A_v,\Pi]$ only has component along the broken generator.

At this stage we need to specify the interpolating field we use for the heavy and the light gauge fields.
We assume that the SM fermion $\psi$  is charged only under one of the gauged $SU(2)$'s, say $SU(2)_1$, which
is the case for models without T-parity.\footnote{For models with T-parity the SM fermion could be charged under both
gauged $SU(2)$'s. However, in this case the heavy gauge boson is T-odd and no tree level effect is generated after integrating out the heavy boson.}
Therefore we will choose $A_1^\alpha$, which
couples to the SM fermionic current, as the interpolating field for the low-energy gauge boson. On the other
hand, it is most convenient to choose the heavy mass eigenstate, $A_h^\alpha=(g_1 A_1^\alpha-g_2 A_2^\alpha)/\sqrt{g_1^2+g_2^2}$, as the heavy field to integrate out, which is linearly independent of (although not orthogonal to) $A_1^\alpha$.
We will be using the following definitions for the gauge couplings
\begin{equation}
g=  \sqrt{g_1^2+g_2^2},\quad \quad g_0 = \frac{g_1 g_2}{\sqrt{g_1^2+g_2^2}},
\end{equation}
where $g$ plays the role of $g_\rho$ and characterizes the mass of the heavy gauge boson $m_{A_h}^2=  g^2 f^2/4$, as we will soon see, and $g_0$ is the gauge coupling for the unbroken gauge 
group.

Given the above digression, we can write
\beq
A_v^\alpha =  g_1 A_1^\alpha - \frac12 g  A_h^\alpha, \quad A_A^\alpha =  \frac12 g A_h^\alpha.
\eeq
Then the leading two-derivative interaction ${\cal L}_{KE}$ in the nl$\sigma$m contains
\bea
\label{eq:jpiva}
{\cal L}_{KE}&=&\frac{f^2}2 {\cal D}^a{\cal D}^a =\frac14 g^2 f^2 A_h^2 +\frac12 D_1\Pi^a D_1\Pi^a 
  -\frac12 g A_h^{\alpha} J_{\Pi v}^{\alpha} + \frac14 g A_h^{\alpha} J_{\Pi A}^{\alpha}+\cdots ,\\
J_{\Pi v}^{\alpha}&=&    i {\rm Tr}(Q_v^{\alpha}[\Pi, D_1\Pi]),\qquad 
 J_{\Pi A}^{\alpha} =   i {\rm Tr}(Q_A^{\alpha}[\Pi, D_1\Pi]) ,
\eea
where $D_1 \Pi =\partial \Pi +i g_1 A_1^\alpha [Q_v^\alpha, \Pi]$ is the gauge-covariant derivative of 
$\Pi$ with respect to the low-energy field $A_1$. We see explicitly that the heavy gauge boson mass
is $m_{A_h}^2= g^2f^2/2$.\footnote{In our normalization ${\rm Tr}(Q_v^\alpha Q_v^\beta)={\rm Tr}(Q_A^\alpha Q_A^\beta)=2\delta^{\alpha\beta}$.} Notice that
in the case of symmetric coset, $J_{\Pi A}^{\alpha}=0$. Since we are only interested in effects of
integrating out $A_h$ at the tree level, which entirely come from the cubic interaction of $A_h$
with two Goldstone fields, we have dropped quadratic interactions in Eq.~(\ref{eq:jpiva}).
In addition to the Goldstone current coupling to $A_h$, we also need to work out the
Yang-Mills Lagrangian in the basis $(A_1, A_h)$:
\begin{eqnarray}
{\cal L}_{YM} &=& -\frac14 (F_{1\, \mu\nu}^\alpha)^2 - \frac14 (F_{2\, \mu\nu}^\alpha)^2  \nonumber \\
  &=& -\frac{1}{4g_0^2} (F_{1\, \mu\nu}^\alpha)^2 -\frac{g^2}{4g_2^2}(H_{\mu\nu}^\alpha)^2 - \frac{g}{g_2^2} J_{F\,\mu}^\alpha A_h^{\alpha\,\mu}  \nonumber\\
  &&\quad -
  \frac{g^2}{2g_2^2} f^{\alpha\beta\gamma} F_{1\,\mu\nu}^\alpha A_{h\,\mu}^\beta A_{h\,\nu}^\gamma + {\cal O}(A_h^3) ,
  \end{eqnarray}
where we have rescaled $A_1\to A_1/g_1$ and the rank-two tensor $H_{\mu\nu}^a$ is defined as $H_{\mu\nu}^\alpha \equiv
 D_{1\,\mu}A_{h\,\nu}^\alpha-D_{1\,\nu}A_{h\,\mu}^\alpha$. 
  We have also defined the gauge current 
 \begin{equation}
 J_{F}^{\alpha\,\mu} = (D_{1\,\nu}F_1^{\nu\mu})^\alpha .
 \end{equation}
From the Lagrangian ${\cal L}_{KE}+{\cal L}_{YM}$ we see the relevant interactions for integrating out $A_h$
at the tree level is
\bea
{\cal L}_{KE}+{\cal L}_{YM} &\supset& \frac12 m_{A_h}^2 (A_h)^2 + A_h^\alpha J_{tot}^\alpha\ ,\\
\label{eq:gaugecurrent}
 J_{tot}^\alpha& =& -\frac12 g J_{\Pi_v}^\alpha +\frac14 g J_{\Pi_A}^\alpha-\frac{g}{g_2^2}  J_{F}^\alpha .
\eea
Integrating out $A_h$ then generates the following dimension-six operators:
\begin{equation}
-\frac1{2m_{A_h}^2} (J_{tot}^\alpha)^2=- \frac1{4f^2}\left(J_{\Pi_v}-\frac12 J_{\Pi_A}\right)^2 -\frac{1}{f^2}  \frac{1}{g_2^2} \left(J_{\Pi_v}-\frac12 J_{\Pi_A}\right) J_F -  \frac1{f^2} \frac{1}{g_2^4}  J_F^2,
 \end{equation}
where the first term $J_\Pi^2$ contains ${\cal O}_H$ and ${\cal O}_r$, the second term gives the operator proportional to $c_W$ and $c_B$, while the last term is the operator ${\cal O}_{2W}$. The important observation here, which we use later to establish the positivity of $c_H$ in little Higgs models, is the coefficients of ${\cal O}_H$ and ${\cal O}_r$
do not depend on the gauge coupling constant.

To compute the coefficients $c_H$ and $c_y$ explicitly,
it is convenient to define the following
fourth-rank tensor that is very similar to ${\cal T}^{abcd}$ in Eq.~(\ref{t1}):
\beq
\overline{\cal T}^{abcd}
= \frac1{4}{\rm Tr}\left\{\left( Q_v^\alpha - \frac12 Q_A^\alpha\right) 
[X^a, X^c]\right\}  \times {\rm Tr}\left\{\left(Q_v^\alpha - \frac12 Q_A^\alpha\right) 
[X^b, X^d]\right\} .
\eeq
Then 
\beq
-\frac{1}{4f^2}\left[ J_{\Pi_v}^\alpha - \frac12 J_{\Pi_A}^\alpha\right]^2 
  \supset \frac1{f^2}h^a h^b \partial_\mu h^c \partial^\mu h^d \overline{\cal T}^{abcd}.
\eeq 
$\overline{\cal T}^{abcd}$ is similar to ${\cal T}^{abcd}$ because $(ac)$ and $(bd)$ components are 
anti-symmetric, and the same reasoning leading to Eq.~(\ref{eq:nlsmT}) implies
\begin{equation}
\overline{\cal T}^{abcd} = \bar{\alpha}_+ (\delta^{ab}\delta^{cd}-\delta^{ad}\delta^{bc}) +\frac{\bar{\beta}}4
 E^{ac} E^{bd}. 
\end{equation}
The sign of the coefficient $\bar{\alpha}_+$ can be deduced from looking at, for example, the 
$\overline{T}^{aacc}$ component:
\beq
\bar{\alpha}_+=\overline{T}^{aacc}= -\frac12 \sum_I\left[  f^{aci} {\rm Tr}(Q_v^\alpha T^i)
        -\frac12 f^{ace} {\rm Tr}(Q_A^\alpha X^e)\right]^2 < 0.
\eeq
Thus we have, identical to Eq.~(\ref{eq:chnlsm}),
\beq
\label{eq:chvec}
c_H^{(v)} = -6 \bar{\alpha}_+ >0, \quad c_y^{(v)}= 2 \bar{\alpha}_+ <0, \quad
   c_H^{(v)}+2c_y^{(v)} = -2\bar{\alpha}_+ >0.
\eeq
where the signs of $c_H$ and $c_H+2c_y$ are both positive.


\end{document}